\documentclass[fleqn]{article}
\usepackage{fontspec}
\usepackage[nobottomtitles*]{titlesec}

\usepackage{float}
\usepackage{needspace}
\usepackage{booktabs}
\usepackage{graphicx}
\usepackage{xcolor}
\usepackage{amsmath,amssymb}
\usepackage{pifont}
\usepackage{textcomp}
\usepackage{array}
\usepackage{enumitem}
\setlist[itemize,1]{leftmargin=1.25em, itemindent=0pt, topsep=3pt, parsep=0pt}
\usepackage{newunicodechar}
\newunicodechar{∈}{\ensuremath{\in}}
\newunicodechar{⊕}{\ensuremath{\oplus}}
\newunicodechar{↦}{\ensuremath{\mapsto}}
\newunicodechar{∎}{\ensuremath{\blacksquare}}
\newunicodechar{⌈}{\ensuremath{\lceil}}
\newunicodechar{⌉}{\ensuremath{\rceil}}
\newunicodechar{⌊}{\ensuremath{\lfloor}}
\newunicodechar{⌋}{\ensuremath{\rfloor}}
\newunicodechar{①}{\textcircled{1}}
\newunicodechar{②}{\textcircled{2}}
\newunicodechar{③}{\textcircled{3}}
\newunicodechar{ᵢ}{\ensuremath{_i}}
\newunicodechar{∪}{\ensuremath{\cup}}
\newunicodechar{–}{--}
\newunicodechar{—}{---}

\newcommand{\Enfittab}[1]{{\footnotesize\setlength{\extrarowheight}{2pt}\resizebox{\linewidth}{!}{#1}}}

\AtBeginDocument{%
  \setlength{\parindent}{0pt}%
  \setlength{\leftskip}{0pt}%
  \setlength{\parskip}{6pt plus 2pt minus 1pt}%
  \linespread{1.15}\selectfont%
  \clubpenalty=10000 \widowpenalty=10000%
  \raggedbottom%
  \let\mwmsection\section%
  \renewcommand{\section}{\clearpage\mwmsection}%
    \setmainfont{texgyretermes-regular.otf}[BoldFont=texgyretermes-bold.otf,ItalicFont=texgyretermes-italic.otf,BoldItalicFont=texgyretermes-bolditalic.otf,Ligatures=TeX]%
  \setmonofont{texgyrecursor-regular.otf}[BoldFont=texgyrecursor-bold.otf,ItalicFont=texgyrecursor-italic.otf,BoldItalicFont=texgyrecursor-bolditalic.otf]%
  \def\keywordname{{\bfseries \emph{Keywords:}}}%
}

\usepackage{preprint}
\usepackage{hyperref}
\usepackage[numbers,square]{natbib}

\usepackage{amsmath}
\usepackage{booktabs}
\usepackage{graphicx}
\usepackage{url}

\hypersetup{colorlinks=true, linkcolor=purple, urlcolor=blue, citecolor=cyan, anchorcolor=black}

\usepackage{xcolor}
\usepackage{lineno}					
\usepackage{tikz} 					

\usepackage{newfloat}
\DeclareFloatingEnvironment[name={Supplementary Figure}]{suppfigure}
\usepackage{sidecap}
\sidecaptionvpos{figure}{c}

\usepackage{titlesec}
\titlespacing\section{0pt}{12pt plus 3pt minus 3pt}{1pt plus 1pt minus 1pt}
\titlespacing\subsection{0pt}{10pt plus 3pt minus 3pt}{1pt plus 1pt minus 1pt}
\titlespacing\subsubsection{0pt}{8pt plus 3pt minus 3pt}{1pt plus 1pt minus 1pt}

\definecolor{lime}{HTML}{A6CE39}

\title{HSRAI: Permutation-Preserving Address Interleaving with Hierarchical Balance Metrics}

\usepackage{authblk}

\author[1\thanks{\texttt{miaomiao\_tong@outlook.com}}]{Xiaotong Yuan}
\affil[1]{Enrigin Technology Co., Ltd.}

\begin{document}

\maketitle

\begin{abstract}
\linespread{1.5}\selectfont\setlength{\leftskip}{3em}\setlength{\rightskip}{3em}\setlength{\parindent}{0pt}%
Address interleaving balances bandwidth across caches, DRAM, and GPU partitions. In a multi-level interconnect topology, the mapping must be a one-to-one, invertible correspondence between logical and encoded addresses and, under typical access patterns, keep traffic uniform at every level's egress ports, not only at terminal slave nodes.

Using random access as the stimulus and terminal uniformity as the acceptance criterion is insufficient for cascaded interconnects; this paper partitions workloads by access-pattern priority and requires a global bijection with no slave node left unvisited for extended periods. Per-level coefficient of variation (CV), consecutive same-port run length, and sliding-window peak occupancy evaluate traffic at each level's egress.

HSRAI preserves high-order and intra-line low-order address bits and applies a W-bit bijection only to the intermediate index segment. Full-domain topologies use offline GF(2) affine search with matrix and salt parameters selected via priority-ranked access patterns and per-level admissibility hard constraints; pruned topologies combine the Chinese Remainder Theorem with remapping.

Evaluation uses a reproducible C++ benchmark covering linear streams, matrix tiling, and 2D arithmetic lattices. On the 128-node full-domain topology, the proposed affine map satisfies bijection, terminal balance, design-time admissibility thresholds, and zero long-window starvation on the high-priority acceptance set; fixed XOR and folding/table baselines expose intermediate-level hotspots or extended zero-access periods. For pruned topologies, the CRT variant significantly improves per-level metrics on linear and tiling accesses. Artifact: \href{https://github.com/xiaotongyuan/hsrai\_address\_hash}{https://github.com/xiaotongyuan/hsrai\_address\_hash} (tag paper-v11).\\
\end{abstract}

\keywords{address interleaving, memory interleaving map, bijective mapping, permutation-preserving, hierarchical balance, affine search, arithmetic lattice access, cascaded crossbar, network-on-chip}

\vspace{0.5cm}


\section{Introduction}

Improving memory-level parallelism has long relied on mixing and interleaving address bits: from XOR-based cache indexing\textsuperscript{[1]} and DRAM bank hashing\textsuperscript{[4]}, to GPU partition balancing\textsuperscript{[5]} and hardware address scrambling, industry has broadly adopted similar techniques at the cache, on-chip scratchpad, DRAM controller, and GPU memory-partition levels. These precedents demonstrate that structured memory accesses create hotspots at memory banks, partitions, or interconnect ports, which is a real problem.

When such interleaving techniques are deployed on multi-level interconnects or cascaded crossbars, the constraints are stronger than for flat bank hashing\textsuperscript{[1,2]}. First, the logical address and the encoded physical address must be in one-to-one, invertible correspondence; it is not permissible to map multiple logical addresses to the same encoded word. An ordinary CRC\textsuperscript{[15]} or a many-to-one hash used solely for bucket assignment cannot directly fulfill this role. Second, the interconnect is hierarchical: the same radix parameter set $\{r_0,r_1,r_2,\ldots\}$ describes the branch width of each crossbar level, and the decoded index corresponds to the egress port number at each level. System bandwidth is typically limited by the most congested port at any level, so optimizing only terminal slave-node uniformity is insufficient; it is also necessary to check intermediate-level load and consecutive same-port accesses using per-level metrics.

In recent years, workloads such as matrix multiplication, attention, and key-value caches have generated a large volume of two-dimensional arithmetic lattice memory accesses (what the memory-bank-conflict literature calls multi-stride access patterns), in which addresses vary as a linear combination of two loop indices. Fixed XOR or folding/table mappings are prone to periodic resonance with such patterns, leaving some slaves unvisited for extended periods. Relying solely on random stimuli and terminal-node count uniformity makes intermediate-level imbalance difficult to expose. Later experiments compare simplified variants of the XOR/affine\textsuperscript{[2]} algorithms from the published literature, search-based mapping approaches, and an internally tuned folding implementation.

This paper studies permutation-preserving address interleaving on multi-level interconnect fabrics. Compared with classical flat bank hashing, the present work additionally emphasizes three requirements: the encoding must be invertible, not merely extracting some bits as a many-to-one bank number; acceptance must cover each level's egress ports across the interconnect, not just terminal slave-node counts; parameter selection must impose hard constraints according to access-pattern priority, and must optimize low-priority metrics within the feasible set. These three points have not been systematically addressed in the existing cache/DRAM/GPU interleaving literature.

\needspace{7\baselineskip}\par\addvspace{12pt plus 3pt minus 2pt}\noindent{\large \textbf{Main contributions:}}\par\addvspace{5pt}\noindent

\begin{itemize}
\item We formulate the multi-level address interleaving problem as a reversible, permutation-preserving mapping, with an invertible address segmentation rule and bijectivity proof (\S{}2.1).
\item Per-level balance metrics, consecutive same-port run length, and sliding-window peak occupancy are introduced as static acceptance measures, showing that terminal uniformity alone or random traffic alone is insufficient to characterize cascaded interconnect risk (\S{}5).
\item The HSRAI scheme is proposed: full-domain topologies use GF(2) affine transforms with priority constraints (\texttt{affine\_pri}); pruned topologies use CRT\textsuperscript{[8--10]}/remap transforms (\texttt{hsrai\_crt474}/\texttt{hsrai\_remap}); core matrices and parameters are generated automatically by a priority-driven search (\S{}4). HSRAI builds on established algebraic components: GF(2) affine\textsuperscript{[2]}, CRC salt\textsuperscript{[15]}, and CRT\textsuperscript{[8--10]}. The main contribution lies in the problem definition framework (permutation-preserving per-level acceptance) and evaluation methodology (priority admissibility + per-level metrics).
\item A reproducible C++ benchmark with structured synthetic access patterns is provided; simplified variants from the published literature are compared quantitatively against HSRAI. Evaluation boundaries and negative results (256-node retarget, CRT factor-7 degradation, etc.) are in \S{}8.2--\S{}8.4; broader scope limitations are discussed in Appendix A.
\end{itemize}

\section{Problem Definition and Existing Approaches}

Industry and academia have taken three broad approaches to address interleaving: approach A (folding/table), approach B (GF(2)/XOR/affine), and approach C (number-theoretic/CRT/remap), each with its own strengths and weaknesses (\S{}2.2). After stating the three constraints: global bijection, multi-level acceptance, and priority-driven search, this paper analyzes the limitations of each approach and proposes HSRAI on that basis.

\needspace{10\baselineskip}
\subsection{Topology, Routing Fields, and Address Bijection}

\needspace{7\baselineskip}\par\addvspace{12pt plus 3pt minus 2pt}\noindent{\large \textbf{Mixed-radix routing fields}}\par\addvspace{5pt}\noindent

The radix $\{r_0,r_1,r_2,\ldots\}$ describes the number of output ports at each level of the cascaded crossbar, from outermost to innermost. During routing, the terminal slave-node index can be viewed as a mixed-radix expansion that nests the radix levels from outermost to innermost, yielding unique per-level routing fields $\{f_0,f_1,\ldots\}$: $f_l$ (innermost) corresponds to the terminal slave node, and higher-level fields $f_0$, $f_1$, etc. correspond to intermediate egress port numbers. Taking $\{r_0{=}4, r_1{=}8, r_2{=}4\}$ (128 slave nodes) as an example, routing fields $\{f_0,f_1,f_2\}$ satisfy $f_i \in [0, r_i)$:
\begin{align}
&\text{slave idx} = f_0 \cdot r_1 \cdot r_2 + f_1 \cdot r_2 + f_2 \nonumber \\
&\text{(L0–L1 combined egress index)} = f_0 \cdot r_1 + f_1 \nonumber \\
&\text{level-0 egress number (L0, outermost)} = f_0
\end{align}

Figure 0 illustrates the radix \texttt{\{4,8,4\}} cascaded crossbar topology (128 slave nodes).

\includegraphics[width=\linewidth]{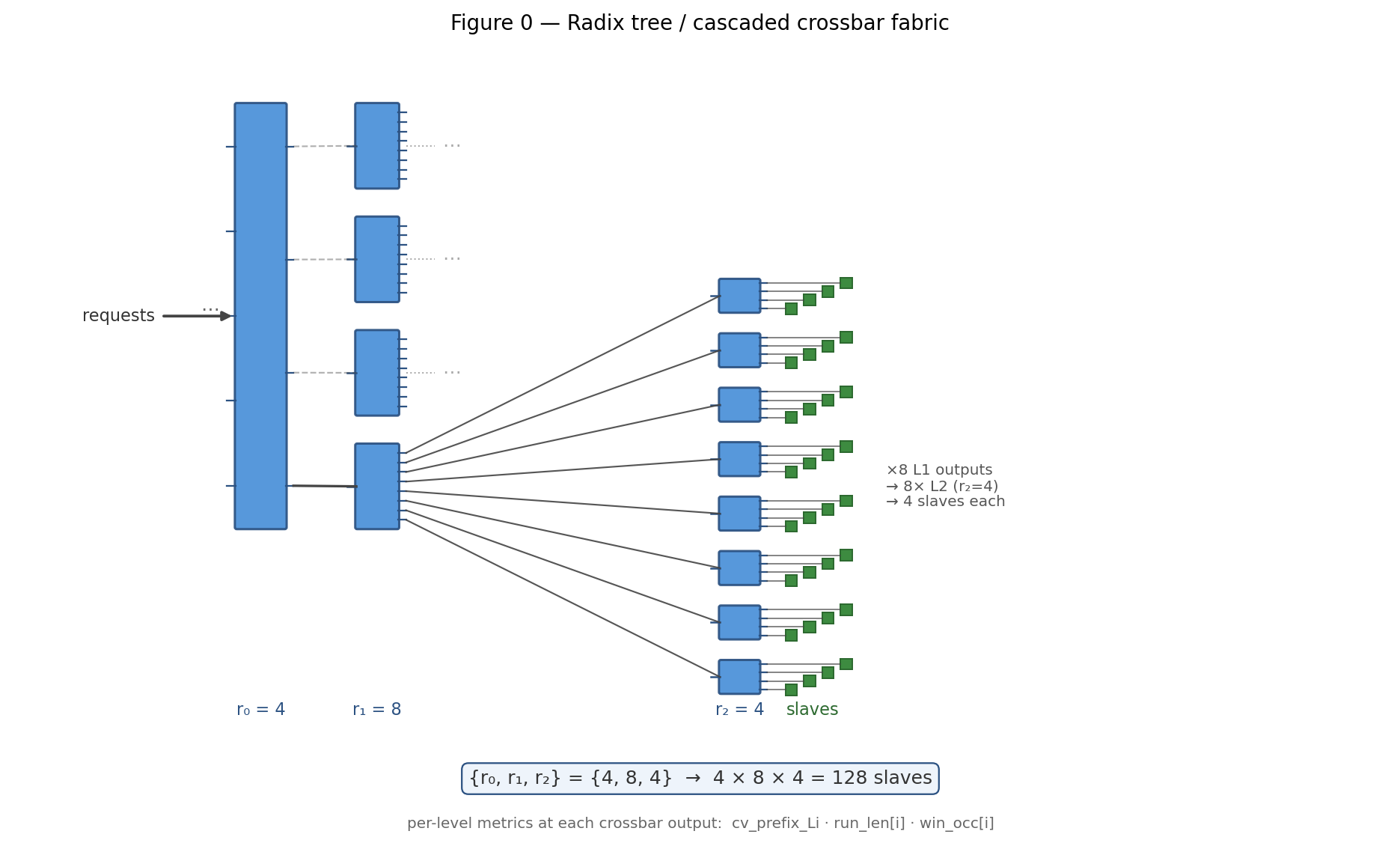}

\needspace{7\baselineskip}\par\addvspace{12pt plus 3pt minus 2pt}\noindent{\large \textbf{low / raw / high layout}}\par\addvspace{5pt}\noindent

Assume the address word accommodates three fields: an intra-line offset, an intermediate index segment, and high-order bits. With minimum access granularity (line size) $L = \log_2(\texttt{line\_bytes})$ bits and intermediate index width $W = \lceil\log_2(\text{slave count})\rceil$:
\begin{align}
&\text{low} = A[L-1:0] \nonumber \\
&\text{raw} = A[L+W-1:L] \nonumber \\
&\text{high} = A[H-1:L+W]
\end{align}
where \texttt{H} is the total address bit width.

\begin{itemize}
\item low: Intra-line byte offset. Does not participate in address interleaving; preserves cache-line / burst uniqueness.
\item raw: The only \texttt{W}-bit segment that is permuted. Its domain in full-domain topologies is $[0, 2^W)$.
\item high: Preserved unchanged. Used only as the source of the permutation parameter (salt), so that different address pages or line blocks use different permutation instances.
\end{itemize}

For full-domain power-of-two topologies, the encoded word is packed as:
\begin{align}
&E = (\text{high} \mathbin{<<} (L+W)) \;|\; (\text{idx} \mathbin{<<} L) \;|\; \text{low}
\end{align}
where \texttt{idx = Permute(raw, salt(high))}. Because \texttt{high} and \texttt{low} do not participate in the permutation, as long as \texttt{raw} $\ensuremath{\mapsto}$ \texttt{idx} is a permutation over $[0, 2^W)$ for a fixed \texttt{high}, the overall logical-to-encoded address mapping is bijective.

\needspace{7\baselineskip}\par\addvspace{12pt plus 3pt minus 2pt}\noindent{\large \textbf{Pruned topologies and harvest mixed-radix compaction}}\par\addvspace{5pt}\noindent

When the number of active slave nodes \texttt{N\_active} is not a power of two (for example, symmetric pruning \texttt{\{4,7,4\}} gives 112 nodes), simply truncating \texttt{raw} to a narrower bit width loses invertibility. In engineering practice, harvest mixed-radix compaction is applied first: surviving nodes are renumbered into \texttt{[0, N\_active)} according to the actual radix, and the permutation is then applied to this compact domain. A typical packing is:
\begin{align}
&\text{line} = A \mathbin{>>} L \nonumber \\
&r = \text{line} \bmod N_{\text{active}} \nonumber \\
&\text{high}' = \lfloor \text{line} / N_{\text{active}} \rfloor \nonumber \\
&\text{idx} = \text{Permute}_{N_{\text{active}}}(r, \text{salt}(\text{high}')) \nonumber \\
&E = ((\text{high}' \cdot N_{\text{active}} + \text{idx}) \mathbin{<<} L) \;|\; \text{low}
\end{align}
As long as \texttt{Permute\_\{N\_active\}} is a permutation over \texttt{[0,N\_active)}, the pairs \texttt{(high', r)} and \texttt{(high', idx)} correspond one-to-one, and the overall address mapping remains bijective.

\needspace{7\baselineskip}\par\addvspace{12pt plus 3pt minus 2pt}\noindent{\large \textbf{Routing field decoding}}\par\addvspace{5pt}\noindent

Given an encoded address \texttt{E}, first unpack to obtain \texttt{(high, idx, low)}, then decode \texttt{idx} into $\{f_0,f_1,\ldots\}$. For full-domain topologies (all $r_i$ are powers of two), this is done by bit-slicing, which is pure wiring in hardware with no arithmetic (e.g., for \texttt{\{4,8,4\}}: the top 2 bits are $f_0$, the middle 3 bits are $f_1$, the bottom 2 bits are $f_2$). For pruned topologies (containing non-power-of-two radix values, such as \texttt{\{4,7,4\}}), modulo operations are applied level-by-level: $f_2 = \text{idx} \bmod r_2$, $f_1 = \lfloor\text{idx}/r_2\rfloor \bmod r_1$, $f_0 = \lfloor\text{idx}/(r_1 \cdot r_2)\rfloor$; equivalently, a CRT construction\textsuperscript{[8--10]} can be used (expanded in \S{}4.3).

\needspace{7\baselineskip}\par\addvspace{12pt plus 3pt minus 2pt}\noindent{\large \textbf{Full-domain vs. pruned (summary)}}\par\addvspace{5pt}\noindent

\begin{itemize}
\item Full-domain: Slave count $N = 2^W$; \texttt{raw} and \texttt{idx} share width \texttt{W}; decoding is pure wiring. Examples: full-domain 128 slave nodes (\texttt{W=7}) or 256 slave nodes (\texttt{W=8}). Approach B can select a full-rank matrix from \texttt{GL(W,2)} for the intermediate segment permutation.
\item Pruned (harvest): The number of surviving nodes under the physical radix is \texttt{N\_active}, which is not a power of two. Example: symmetric \texttt{\{4,7,4\}} yields 112 active slave nodes. Logical line numbers must first be compacted into \texttt{[0,N\_active)} before permutation. Directly applying a full-domain 128-entry table followed by mod 112 would break bijectivity or introduce ``empty-node'' statistical bias.
\item Engineering implication: When changing topologies, one must first determine whether the topology is full-domain or pruned, identify \texttt{N\_active} and the compaction rule, then select the hash algorithm.
\end{itemize}

\begin{quote}
The Pack/Unpack formulas above use a general canonical layout for illustration. The actual implementation for 128 slave nodes (\{4,8,4\}) in the experiment code uses a scattered bit-field layout; see \S{}7.1 for implementation differences.
\end{quote}

\needspace{10\baselineskip}
\subsection{Overview and Limitations of Existing Approaches A/B/C}

Methods used in industry and academia to address similar problems can be grouped into three approaches A, B, and C; each falls short under the bijection, hierarchical-topology, and parameterizability requirements.

\needspace{7\baselineskip}\par\addvspace{12pt plus 3pt minus 2pt}\noindent{\large \textbf{Approach A: folding / table}}\par\addvspace{5pt}\noindent

Representative idea: map middle and high-order address segments to routing fields by bit folding (integer addition); multiple permutation tables can be cascaded while preserving the \texttt{high} bits and maintaining a \texttt{raw} $\ensuremath{\mapsto}$ \texttt{idx} bijection. This approach is common in industrial implementations, but there is no publicly available benchmark that directly aligns with the ``global bijection + per-level acceptance'' setting of this paper.

Advantages:

\begin{itemize}
\item Under fixed topology and fixed access patterns, manual tuning can achieve very low CV. For example, CV of access counts can be 0 for common patterns such as sequential streams, power-of-two strides, and matrix tiling.
\item The table is a permutation \ensuremath{\rightarrow} bijection is naturally satisfied, with no additional decoding logic.
\end{itemize}

Limitations:

\begin{itemize}
\item The periodic structure of the folding function is prone to resonance with structured strides: for example, with 128 slave nodes \texttt{\{4,8,4\}}, a linear sweep with stride $(N -1) \times \text{line\_bytes}$ (stride 127 in line-index units when $N=128$) aligns the raw index progression with the folding period. In the stride-127 long-window test (\texttt{make stride127}, \S{}8.1.2), the internal \texttt{type8} reference has 33/128 slave nodes with zero accesses over extended windows. Once an access pattern falls outside the tuning range, it is difficult to remedy by minor parameter adjustment.
\end{itemize}

\needspace{7\baselineskip}\par\addvspace{12pt plus 3pt minus 2pt}\noindent{\large \textbf{Approach B: GF(2) / XOR / affine}}\par\addvspace{5pt}\noindent

Representative references: Vandierendonck \& De Bosschere XOR bank hash\textsuperscript{[2]}; configurable XOR matrix for GPU scratchpad\textsuperscript{[5]}; Minimalist Open-page DRAM bank XOR\textsuperscript{[4]}; skewed-associative cache\textsuperscript{[3]}.

Advantages:

\begin{itemize}
\item Low hardware cost: a fixed \texttt{W}\,\ensuremath{\times}\,\texttt{W} binary matrix-vector multiply with XOR, no large ROM.
\item A full-rank matrix is invertible over GF(2) \ensuremath{\rightarrow} for fixed \texttt{high}, \texttt{raw} $\ensuremath{\mapsto}$ \texttt{idx} is a permutation, satisfying bijection.
\item Naturally suited to $2^n$ full-domain canonical bit-slicing; parameters \texttt{(A, salt)} can be generated automatically by topology-driven search, without manually computing permutation tables.
\end{itemize}

Limitations:

\begin{itemize}
\item Full rank guarantees only that \texttt{raw} $\ensuremath{\mapsto}$ \texttt{idx} is a permutation; it does not ensure that per-level CV/run/win constraints are met. For example, using a near-identity matrix (\texttt{generic\_affine}) with stride-4 sweep: $f_0 \approx \texttt{raw[6:5]}$ changes only every 8 cycles (theoretical L0 run length up to 8), so the outermost (L0) port sees consecutive hits. Although terminal \texttt{cv\_slave} = 0, there is already a structural hotspot at the outer level. Mapping $f_0$ to the low-order bits of \texttt{raw} is not a robust fix either: it improves L0 rotation only for unit-stride (consecutive-line) accesses, and other small strides that leave those bits unchanged can still bypass outer-level port rotation.
\item Plain many-to-one CRC/LFSR hash\textsuperscript{[15]} does not satisfy address bijection and is outside the scope of this work.
\end{itemize}

\needspace{7\baselineskip}\par\addvspace{12pt plus 3pt minus 2pt}\noindent{\large \textbf{Approach C: number-theoretic / CRT / prime modulo / remap}}\par\addvspace{5pt}\noindent

Representative references: Odd Memory Systems\textsuperscript{[8]}; Gao CRT prime memory\textsuperscript{[9]}; Valero modulo skewing\textsuperscript{[10]}.

Advantages:

\begin{itemize}
\item When \texttt{N\_active} is not a power of 2, especially when it factors as $2^a \times \text{small odd number}$, CRT can construct permutation matrices on each factor domain separately and then combine them; suitable for harvest topologies.
\item Naturally breaks resonance between pure power-of-two stride accesses and bit-slicing (odd modulus approach).
\end{itemize}

Limitations:

\begin{itemize}
\item In full-domain $2^n$ designs, the timing and area cost of CRT/division is generally less clean than GF(2) affine: for radix \texttt{\{4,8,4\}}, $W=7$, GF(2) affine needs only 7 XOR trees; the CRT path additionally requires an odd-modulus divider, making it unsuitable as the first choice for full-domain topologies.
\item Remap / cycle-walk algorithms guarantee bijection for arbitrary \texttt{N\_active}, but may have resonance for certain access shapes, and in the worst case require multiple iterations to obtain a valid idx, which requires a multi-cycle RTL pipeline.
\end{itemize}

\needspace{7\baselineskip}\par\addvspace{12pt plus 3pt minus 2pt}\noindent{\large \textbf{Relationship of the three approaches to this problem}}\par\addvspace{5pt}\noindent

\needspace{20\baselineskip}
\textbf{Table 1} --- A/B/C approaches vs. problem constraints:

\bigskip\noindent
\par\noindent\Enfittab{\begin{tabular}{>{\raggedright\arraybackslash}m{\dimexpr 0.250\linewidth-2\tabcolsep}>{\raggedright\arraybackslash}m{\dimexpr 0.250\linewidth-2\tabcolsep}>{\raggedright\arraybackslash}m{\dimexpr 0.250\linewidth-2\tabcolsep}>{\raggedright\arraybackslash}m{\dimexpr 0.250\linewidth-2\tabcolsep}}
\toprule
Constraint & A folding/table & B GF(2) affine & C CRT/remap \\
\midrule
Address bijection & Yes (permutation table) & Yes (full-rank A) & Yes (finite-domain permutation) \\
Per-level CV / window (\S{}5) & By manual table tuning & Needs added level-aware search & Requires dedicated construction \\
Topology parameterizability & Weak & Strong & Medium \\
Arithmetic lattice cliff & Easy to encounter & Can be mitigated & Possible \\
\bottomrule
\end{tabular}}

\bigskip

Summary: Approach A is good at manual tuning for fixed topologies, but lacks parameterization and is prone to cliffs under lattice-type stimuli. Approach B suits full-domain $2^n$ scenarios with low hardware cost, but without level-aware search it is hard to meet per-level CV/window targets. Approach C suits harvest pruned scenarios; its hardware cost is higher in full-domain topologies.

\section{Basic Mathematical Properties}

This section gives several properties shared by bijective address interleaving schemes, with proofs.

\par\addvspace{4pt}\noindent\rule{\linewidth}{0.4pt}\par\addvspace{2pt}\noindent \textbf{Property 1} (Address bijection in full-domain topologies)\par\addvspace{10pt}\noindent
Let the slave count be $2^W$, with the \texttt{high} and \texttt{low} field layout fixed (their values vary across addresses but are held unchanged by the transform). If, for each \texttt{high}, $P_{\text{high}}$ is a permutation over $[0, 2^W)$, then
\begin{align}
&\text{idx} = P_{\text{high}}(\text{raw}) \nonumber \\
&E = \text{Pack}(\text{high},\; \text{idx},\; \text{low})
\end{align}
is a bijection from logical to encoded addresses. The inverse is $\text{raw} = P_{\text{high}}^{ -1}(\text{idx})$ followed by unpack.

\textbf{Proof:} For any fixed $(\text{high},\text{low})$, $P_{\text{high}}$ is a permutation over $[0,2^W)$, so $\text{raw} \ensuremath{\leftrightarrow} \text{idx}$ is one-to-one; \texttt{Pack} is a bijection under fixed fields (\S{}2.1), so within a slice $\text{raw} \ensuremath{\leftrightarrow} E$ is bijective. Moreover, \texttt{Pack} preserves the \texttt{high} and \texttt{low} bits verbatim, so the $E$-images of different $(\text{high},\text{low})$ slices are pairwise disjoint; the union of disjoint bijections is a global bijection. \ensuremath{\blacksquare}

For the affine instance $\text{idx} = A \cdot \text{raw} \oplus \text{salt}(\text{high})$, the map $\text{raw} \ensuremath{\mapsto} \text{idx}$ is a permutation if and only if $A$ is full-rank over $\text{GF}(2)$ (Property 2).

\par\addvspace{4pt}\noindent\rule{\linewidth}{0.4pt}\par\addvspace{2pt}\noindent \textbf{Property 2} (Full-rank GF(2) affine maps are permutations)\par\addvspace{10pt}\noindent
Let $A \in \text{GF}(2)^{W \times W}$ be full-rank and $c \in \text{GF}(2)^W$ arbitrary. The map
\begin{align}
&\varphi(\text{raw}) = A \cdot \text{raw} \oplus c
\end{align}
is a bijection over $[0, 2^W)$. The constant $c$ (corresponding to salt) is just an XOR offset and does not affect invertibility.

\textbf{Proof:} Full rank $\ensuremath{\Rightarrow}$ $A$ is invertible over $\text{GF}(2)$. If $\varphi(\text{raw}_1) = \varphi(\text{raw}_2)$, then
\begin{align}
&A \cdot (\text{raw}_1 \oplus \text{raw}_2) = 0 \implies \text{raw}_1 = \text{raw}_2
\end{align}
(injection). Domain and range are both finite sets of $2^W$ elements; an injection on a finite set is also a surjection, hence a permutation (by the pigeonhole principle). Conversely, if $A$ is not full-rank, there exists $\text{raw}_0 \neq 0$ with $A \cdot \text{raw}_0 = 0$, so $\varphi(\text{raw}_0) = \varphi(0) = c$, not injective; full rank is therefore also necessary. \ensuremath{\blacksquare}

\par\addvspace{4pt}\noindent\rule{\linewidth}{0.4pt}\par\addvspace{2pt}\noindent \textbf{Property 3} (Exactly uniform terminal slave access over a complete raw block)\par\addvspace{10pt}\noindent
Fix \texttt{high}. If \texttt{raw} ranges over $[0, 2^W)$ exactly once, then for any permutation $P_{\text{high}}$, \texttt{idx} also takes each value exactly once; provided the terminal slave is indexed directly by \texttt{idx} ($M_L = I$, slave count $= 2^W$), $\text{cv\_slave}$ (\S{}5) $= 0$ over that block.

\textbf{Proof:} A permutation does not change the multiset $\{P_{\text{high}}(\text{raw}) : \text{raw} \in [0, 2^W)\}$; as \texttt{raw} ranges over $[0, 2^W)$ once, each \texttt{idx} appears exactly once, so terminal slave-node counts are exactly uniform. \ensuremath{\blacksquare}

This property shows that permutation-preservation is a necessary foundation, but it alone does not guarantee:

\begin{itemize}
\item Sequential uniformity within short windows (\texttt{run\_len} (\S{}5) may be large);
\item Uniformity of the level-$i$ projections $f_i(\text{idx})$ (\texttt{cv\_prefix\_Li} (\S{}5) may be $> 0$);
\item In practice, experiment windows rarely cover the complete raw domain, so nonzero \texttt{cv\_slave} is a normal finite-window effect (\S{}7.1).
\end{itemize}

For pruned topologies ($N_\text{active} \neq 2^W$), the uniformity claim requires $P_\text{high}$ to be a permutation over the compact domain $[0, N_\text{active})$ and \texttt{cv\_slave} to be computed over that same domain; see Property 5.

\needspace{14\baselineskip}\par\addvspace{4pt}\noindent\rule{\linewidth}{0.4pt}\par\addvspace{2pt}\noindent \textbf{Property 4} (Algebraic criterion for per-level camping in GF(2) affine maps)\par\addvspace{10pt}\noindent
\textit{Definition:} Two accesses are said to camp at level $l$ if they are routed to the same port at that level (i.e., their $f_l$ values coincide). Sustained camping saturates the throughput of that egress port and causes local congestion (cf. GPU partition camping\textsuperscript{[7]}).

Let $\text{idx} = A \cdot \text{raw} \oplus \text{salt}(\text{high})$ ($A$ full-rank; for fixed \texttt{high}, $\text{raw} \ensuremath{\leftrightarrow} \text{idx}$ is bijective, by Property 2). The level-$l$ egress is given by projection matrix $M_l \in \text{GF}(2)^{k_l \times W}$ as $f_l = M_l \cdot \text{idx} \in \text{GF}(2)^{k_l}$ ($k_l$ is the port width at that level). Let $d = \text{raw}_1 \oplus \text{raw}_2 \in \text{GF}(2)^W$ (the GF(2) address difference).

\textbf{Case 1 (same high, same salt):} Two addresses camp at level $l$ if and only if
\begin{align}
&M_l \cdot A \cdot d = 0
\end{align}
\textbf{Case 2 (different high):} Let $\Delta\text{salt} = \text{salt}(\text{high}_1) \oplus \text{salt}(\text{high}_2)$. The camping criterion is
\begin{align}
&M_l \cdot (A \cdot d \oplus \Delta\text{salt}) = 0
\end{align}

$\text{salt}(\text{high})$ varies with the high-order bits to break synchronized clustering across \texttt{high} groups. In Case 1 the set of camping differences is $\ker(M_l A)$, a subspace of $\text{GF}(2)^W$; determining which strides camp at level $l$ reduces to finding this subspace.

\textbf{Proof:} With the same \texttt{high}, the salt terms cancel, giving $\text{idx}_1 \oplus \text{idx}_2 = A \cdot d$; camping depends solely on whether $d$ projected through $A$ and $M_l$ is zero. With different \texttt{high}, $f_l^{(1)} = f_l^{(2)}$ if and only if $M_l \cdot (\text{idx}_1 \oplus \text{idx}_2) = 0$; since $\text{idx}_1 \oplus \text{idx}_2 = A \cdot d \oplus \Delta\text{salt}$, this reduces to $M_l \cdot (A \cdot d \oplus \Delta\text{salt}) = 0$. \ensuremath{\blacksquare}

\needspace{7\baselineskip}\par\addvspace{12pt plus 3pt minus 2pt}\noindent{\large \textbf{Usage notes:}}\par\addvspace{5pt}\noindent

\begin{itemize}
\item \textit{Intermediate-level camping does not imply terminal address collision:} A full-rank $A$ guarantees that different \texttt{raw} values within the same \texttt{high} do not map to the same slave (Property 2). Intermediate-level camping affects only that level's throughput and local congestion, not address correctness.
\item \textit{Integer stride and XOR model limitations:} The criterion is exact for a fixed XOR difference $d$. Integer-stride accesses use ordinary addition; the XOR difference between consecutive steps is not constant (carry bits alter individual bit differences), so algebraically predicted camping can be partially broken during a sequential scan. Per-level behavior must be verified empirically using CV/\texttt{run\_len}/\texttt{win\_occ} on structured patterns.
\end{itemize}

\textbf{Engineering implication:} Full-rank $A$ is a hard constraint. The camping algebraic criterion is only a search diagnostic; algorithm selection should be based on measured pattern metrics (\S{}6.2).

\par\addvspace{4pt}\noindent\rule{\linewidth}{0.4pt}\par\addvspace{2pt}\noindent \textbf{Property 5} (Bijection under harvest mixed-radix compaction)\par\addvspace{10pt}\noindent
Let $N_{\text{active}}$ be arbitrary. If $P_{\text{high}}$ is a permutation over $[0, N_{\text{active}})$, then under the pruned packing rule of \S{}2.1, the map $(\text{high}', r) \ensuremath{\mapsto} (\text{high}', \text{idx})$ is a bijection, and the overall address transform is invertible.

\textbf{Proof:} $P_{\text{high}}$ is a permutation over $[0, N_{\text{active}})$, so within a slice $r \ensuremath{\leftrightarrow} \text{idx}$ is one-to-one. The cross-slice global bijection reuses the Pack field-independence argument of Property 1 (the $E$-images of different $(\text{high}', \text{low})$ slices are disjoint), simply replacing the full-domain $[0, 2^W)$ with the compact domain $[0, N_{\text{active}})$. When $N_{\text{active}}$ is not a power of two, $P_{\text{high}}$ is constructed on the surviving set by CRT/remap/cycle-walk, which preserves the permutation property. The Pack rule in \S{}2.1 treats $W$ and $L$ as independent system parameters: $W = \lceil\log_2 N_{\text{active}}\rceil$ is driven by the active slave count, and $L = \log_2(\texttt{line\_bytes})$ by the line size ($L = 7$ here, i.e., bits \texttt{0:6} for a 128 B line; a different line size such as $L = 3$ simply shifts the other fields). When the system changes, both can be re-tuned, with \texttt{high} absorbing the remaining bits so that all routing fields stay consistent with the pruned topology. \ensuremath{\blacksquare}

\section{The HSRAI Scheme}

HSRAI (Hierarchical Stride-Resistant Address Interleaving) is the address interleaving scheme proposed in this paper. It follows the address pack rules of \S{}2.1, applying an invertible permutation to the intermediate segment \texttt{raw} (segment \ensuremath{\rightarrow} permute \ensuremath{\rightarrow} repack), while leaving \texttt{high} and \texttt{low} unchanged. Full-domain topologies (\S{}4.2) use a GF(2) affine map; pruned topologies (\S{}4.3) use CRT or remap. Both share the pipeline of \S{}4.1 and the salt mechanism of \S{}4.4. Key parameters $(A, \text{seed})$ are determined by the offline search described in \S{}4.5 under per-level admissibility constraints. \S{}4.5 gives two delivered instances and Table 3.

\needspace{10\baselineskip}
\subsection{Encode and Decode Pipeline}

Using a 48-bit logical byte address and 128-byte line size ($L=7$, $W=7$) as an example, Encode proceeds in four steps. The formulas below use the canonical layout to illustrate address segmentation and permutation. The scattered bit-field layout used in the 128-slave-node implementation is described in \S{}7.1.

\begin{enumerate}
\item Segment (\S{}2.1):
\begin{align}
&\text{low} = \text{offset}[6:0] \nonumber \\
&\text{raw} = \text{offset}[13:7] \nonumber \\
&\text{high} = \text{offset}[47:14]
\end{align}

\item Compute salt: $\text{salt} \ensuremath{\leftarrow} \text{Salt}(\text{high})$, producing $W$ bits, depending only on high.
\item Permute the intermediate segment:
\begin{align}
&\text{idx} = \text{Permute}(\text{raw},\; \text{salt})
\end{align}

\item Pack: write idx back into the raw field of offset; low and high are unchanged. This gives encoded address $E$. Routing fields $\{f_i\}$ are decoded from idx according to the radix rules of \S{}2.1.
\end{enumerate}

Decode is the reverse: extract \texttt{(high, idx, low)} from $E$, recompute salt from the same high field, execute $\text{raw} = \text{Permute}^{ -1}(\text{idx},\, \text{salt})$, and write raw back to the corresponding field to recover offset.

\needspace{10\baselineskip}
\subsection{Full-Domain Permutation: GF(2) Affine}

For full-domain topologies with $N = 2^W$, Permute expands to an affine map over GF(2):
\begin{align}
&\text{idx} = A \cdot \text{raw} \oplus \text{salt}
\end{align}
where $A$ is a $W \times W$ binary matrix. $A \cdot \text{raw}$ (defined below as the Gf2MatVec function) is a GF(2) matrix-vector product: each output bit is computed by ANDing raw with a row of $A$ and XOR-reducing the result; in hardware this is $W$ XOR trees. Salt is a $W$-bit mask derived from high; its role is described in \S{}4.4.

When $A$ is full-rank, $A \cdot \text{raw}$ is a permutation over $[0, 2^W)$ (\S{}3 Property 2). XORing with salt preserves the permutation property, so for fixed high, raw $\ensuremath{\leftrightarrow}$ idx is bijective.

Decode uses the precomputed $A^{ -1}$:
\begin{align}
&\text{raw} = A^{-1} \cdot (\text{idx} \oplus \text{salt})
\end{align}
The hardware cost is a $W \times W$ XOR network plus salt logic; no $2^W$-entry lookup table is needed.

For $W=7$ and 128 slave nodes, this paper defines two affine configurations.

The first is \texttt{generic\_affine} (baseline comparison): matrix $A_\text{base}$ is close to the identity (the last row is non-trivial). The salt function Salt7FromHigh22 splits the 22-bit high field into a low 20-bit half and an upper 2-bit half (the latter XOR'd with a constant), computes CRC32 over each half separately, XORs the two CRC32 results, and truncates to 7 bits. Parameters are fixed and were not derived by search:
\begin{align}
&\text{idx} = \texttt{Gf2MatVec7}(A_\text{base},\, \text{raw}) \oplus \texttt{Salt7FromHigh22}(\text{high})
\end{align}

The second is \texttt{affine\_pri} (the core algorithm of this paper): same formula structure, but matrix $A_\text{pri}$ and seed are determined by offline search to satisfy hard-constraint metrics under high-priority patterns, then frozen in hardware. The complete formula and search description are in \S{}4.5.

\needspace{10\baselineskip}
\subsection{Pruned Topologies: CRT and Remap}

When $N_\text{active}$ is not a power of two (harvest, \S{}2.1), logical line numbers must first be compacted to $r \in [0, N_\text{active})$ before permutation. Directly applying a full-domain GF(2) affine mod $N_\text{active}$ breaks bijection. HSRAI provides two pruned constructions; theoretical sources include odd-modulus/CRT memory interleaving\textsuperscript{[8--10]}.

\textbf{CRT path} (\texttt{hsrai\_crt474}, $N_\text{active} = 112 = 2^4 \times 7$): Write $N_\text{active} = 2^a \cdot q$ with $q$ odd. Decompose the compact index $r \in [0, 2^a \cdot q)$ by mixed-radix quotient--remainder as $r = s_2 + 2^a \cdot r_q$, where $s_2 = r \bmod 2^a \in [0, 2^a)$ is the coordinate of $r$ in the $2^a$ sub-domain (the low $a$ bits of $r$), and $r_q = \lfloor r / 2^a \rfloor \in [0, q)$ is the coordinate in the odd-factor sub-domain. This decomposition is a bijection $[0, 2^a \cdot q) \ensuremath{\leftrightarrow} [0, 2^a) \times [0, q)$, so $(s_2, r_q)$ is uniquely determined by $r$ and vice versa ($s_2, r_q$ are internal decomposition variables, not to be confused with routing field $f_2$ or radix $r_2$). Apply affine permutations separately on the two sub-domains, then combine with Garner/CRT:
\begin{align}
&p_2 = A_2 \cdot s_2 \oplus \text{salt}_2(\text{high}) \nonumber \\
&p_q = \bigl(a_q \cdot r_q + b_q + \text{salt}_q(\text{high})\bigr) \bmod q
\end{align}
CRT synthesis yields a unique $\text{idx} \in [0, 112)$ satisfying $\text{idx} \equiv p_2 \pmod{2^a}$ and $\text{idx} \equiv p_q \pmod{q}$. Decode reverses each sub-domain permutation and recovers $r$ (bijection guaranteed by Property 5).

\textbf{Remap path} (implementation name \texttt{hsrai\_remap}; referred to as \texttt{hsrai\_cyclewalk474/4742} in benchmarks) is used for two types of pruned topologies: when $N_\text{active}$ cannot be written as $2^a \times q$, or when surviving nodes after harvest are distributed irregularly over the full domain $[0,2^W)$ (surviving set $S$, $|S|=N_\text{active}$). For compact index $r \in [0,N_\text{active})$, Encode computes idx by the following deterministic rule: let $x = r$; apply the full-domain GF(2) affine $x \ensuremath{\leftarrow} A\cdot x \oplus \text{salt}(\text{high})$ repeatedly, up to \texttt{max\_rounds} times; as soon as $x \in S$, set $\text{idx}=x$ and stop. If no element of $S$ is hit after the loop, set $\text{idx}=(c_1 r + c_2)\bmod N_\text{active}$ (where $c_1$ is coprime to $N_\text{active}$). The two paths together define a deterministic encode rule that can be expanded into a lookup table at design time.

At design time, enumerate all $r$ for candidate parameters $(A,\text{seed},S)$ and fill \texttt{enc[r]} according to the rule above. The parameters are only frozen if \texttt{enc} forms a permutation over $[0,N_\text{active})$; the inverse table \texttt{dec[idx]} is then derived. Otherwise the parameter set is discarded. Hardware can store these two tables of $N_\text{active}$ entries each: Encode looks up \texttt{enc[r]}, Decode looks up \texttt{dec[idx]}, with fixed latency per cycle. The public C++ code retains an online-computation version for admissibility evaluation and round-trip self-check. Results for untuned remap baseline instances are in \S{}8.3--\S{}8.4.

Compared to CRT, Remap does not require $N_\text{active}$ to factor as $2^a \times q$ and is more broadly applicable, but requires storing two tables of $N_\text{active}$ entries in hardware, a larger cost than the constant-size GF(2) operations and small modular arithmetic of CRT. CRT should be preferred when the topology is factorable: for example, symmetric harvest reducing full-domain \texttt{\{4,8,4\}} (128 slave nodes) to \texttt{\{4,7,4\}} (112 slave nodes, $N_\text{active}=2^4\times7$) uses \texttt{hsrai\_crt474} (\S{}4.5 instance 2). When the topology cannot be factored, or when surviving nodes after harvest are distributed irregularly (e.g., a global random harvest giving 124 or 120 slave nodes), the remap lookup-table approach is used instead.

\needspace{10\baselineskip}
\subsection{Salt: Role and Computation}

If the permutation is always fixed (salt $\equiv 0$), structured strides will resonate persistently with the radix routing period, repeatedly hitting the same set of egress ports $(f_0, f_1, \ldots)$ (see the camping criterion in Property 4). \texttt{salt(high)} selects a $W$-bit XOR mask for each high block (functionally a tweak in tweakable-cipher terminology): within the block, the same matrix $A$ is used to shuffle the order of raw values, but switching blocks shifts to a different permutation instance, breaking synchronized clustering across pages. \texttt{high} is read-only; in Decode, salt is recomputed from the same high field in the encoded word, requiring no extra storage. A CRC32-truncation implementation can use a configurable GF(2) parallel network\textsuperscript{[15]}.

Common salt functions are listed below (Table 2).

\needspace{18\baselineskip}
\textbf{Table 2} --- Common salt functions:

\bigskip\noindent
\par\noindent\Enfittab{\begin{tabular}{>{\raggedright\arraybackslash}m{\dimexpr 0.333\linewidth-2\tabcolsep}>{\raggedright\arraybackslash}m{\dimexpr 0.333\linewidth-2\tabcolsep}>{\raggedright\arraybackslash}m{\dimexpr 0.333\linewidth-2\tabcolsep}}
\toprule
Salt function & Computation & Parameters \\
\midrule
\texttt{SaltKFromHigh} & \texttt{CRC32(high\_lo) ⊕ CRC32(high\_hi ⊕ 0xA5A5A5A5)}, truncated to $W$ bits & No seed \\
\texttt{SaltKCrcSeed} & \texttt{CRC32(high ⊕ seed)} mod $2^W$ & seed (32-bit) \\
\texttt{SaltKMul} & \texttt{(high $\times$ mult)} multi-level XOR fold and truncate & mult (32-bit odd number) \\
\bottomrule
\end{tabular}}

\bigskip

\needspace{10\baselineskip}
\subsection{Delivered Instances and Parameter Search}

\textbf{Parameter selection procedure (design time):} The parameters $(A, \text{seed})$ for \texttt{affine\_pri} and \texttt{hsrai\_crt474} are obtained by offline search at design time and frozen as constants (Appendix B). The search space is the set of full-rank matrices in $GL(W,2)$ combined with salt configurations (\texttt{salt\_kind}, \texttt{seed}). Candidates that fail the priority admissibility hard thresholds of \S{}6.2 are eliminated first; among feasible candidates, the soft objective $J$ is minimized (primarily optimizing \#5 and non-$2^n$ strides). The public artifact delivers frozen implementations and benchmark evaluation tools (\S{}7.4); the design-time search program is not included. When the topology changes or priorities are adjusted, re-selection must be done in the design environment.

For the main topology (128, \texttt{\{4,8,4\}}), a two-phase random sampling with \texttt{mt19937\_64} (12,000 iterations for the salt phase, 40,000 for the GF(2) matrix phase) found \textbf{17} feasible candidates satisfying all hard constraints of \S{}6.2. The one with the smallest $J$ was frozen as \texttt{affine\_pri}. A subsequent phase-C verification run (40,000 iterations) confirmed that this parameter remains near-optimal under the same objective function, with no better alternative found. The 256-node topology serves as a retarget stress test; results are in \S{}8.2.

\textbf{Instance 1} (full-domain 128 slave nodes, \texttt{hsrai\_affine\_pri}, radix \texttt{\{4,8,4\}}, $W=7$):
\begin{align}
&\text{idx} = \texttt{Gf2MatVec7}(A_\text{pri},\, \text{raw}) \oplus \texttt{Salt7CrcSeed}(\text{high},\; \texttt{0xF1BB3E62})
\end{align}
where Salt7CrcSeed computes $\texttt{CRC32}(\text{high} \oplus \text{seed}) \bmod 2^7$. $A_\text{pri}$ is a 7-row full-rank matrix; seed \texttt{0xF1BB3E62} was selected by offline search after passing hard constraints \#1--\#4, and is frozen in the implementation file. The complete matrix values and topology parameters are in Appendix B.

\textbf{Instance 2} (pruned 112 slave nodes, \texttt{hsrai\_crt474}, radix \texttt{\{4,7,4\}}): following the CRT path of \S{}4.3, $N_\text{active} = 2^4 \times 7 = 112$:
\begin{align}
&p_2 = A_2 \cdot s_2 \oplus \texttt{SaltKCrcSeed}(\text{high},\, \text{seed}_2) \nonumber \\
&p_q = \bigl(a_q \cdot r_q + b_q + \texttt{CRC}(\text{high} \oplus \text{seed}')\bigr) \bmod 7
\end{align}
CRT synthesis yields $\text{idx} \in [0, 112)$; Decode is the reverse. Per-level balance metrics and known degradation scenarios are in \S{}8.3--\S{}8.4 and Appendix A.1.

\needspace{20\baselineskip}
\textbf{Table 3} --- Algorithm quick reference:

\bigskip\noindent
\par\noindent\Enfittab{\begin{tabular}{>{\raggedright\arraybackslash}m{\dimexpr 0.250\linewidth-2\tabcolsep}>{\raggedright\arraybackslash}m{\dimexpr 0.250\linewidth-2\tabcolsep}>{\raggedright\arraybackslash}m{\dimexpr 0.250\linewidth-2\tabcolsep}>{\raggedright\arraybackslash}m{\dimexpr 0.250\linewidth-2\tabcolsep}}
\toprule
Name & Applicable topology & Permutation core & Parameter source \\
\midrule
\texttt{generic\_affine} & Full-domain $N=2^W$ & GF(2) affine $A_\text{base}$ + Salt7FromHigh22 & Fixed, no search; baseline comparison \\
\texttt{affine\_pri} & Full-domain $N=2^W$ & GF(2) affine $A_\text{pri}$ + Salt7CrcSeed(seed) & Search-frozen; core algorithm of this paper \\
\texttt{hsrai\_crt474} & Pruned $N_\text{active}=112$ & CRT factor-wise affine (\S{}4.3) & Search-frozen; preferred for pruned topologies \\
\texttt{hsrai\_remap} & Pruned arbitrary $N_\text{active}$ & GF(2) + cycle-walk\textsuperscript{[18]} (\S{}4.3) & Offline precomputed table; pruned supplement; named \texttt{hsrai\_cyclewalk474} / \texttt{hsrai\_cyclewalk4742} in benchmarks by topology \\
\bottomrule
\end{tabular}}

\bigskip

\section{Per-Level Evaluation Metrics}

This section defines the static acceptance measures evaluated after routing-field decoding (\S{}2.1). CV below refers to the coefficient of variation of access counts (lower is more uniform).

\needspace{7\baselineskip}\par\addvspace{12pt plus 3pt minus 2pt}\noindent{\large \textbf{Design objectives}}\par\addvspace{5pt}\noindent

Admissibility is defined as follows: over the workload stimulus set partitioned by deployment priority, the mapping must simultaneously satisfy three requirements. (i) Global bijection, verified by round-trip consistency checks. (ii) No slave node goes unvisited during extended windows. (iii) Per-level coefficient of variation, consecutive same-port run length, and sliding-window peak occupancy all fall within the hard thresholds set at design time. Admissibility is the design-time acceptance criterion for memory mappings over the full stimulus set of \S{}6.1.

\needspace{18\baselineskip}
\textbf{Table 4} --- Admissibility design objectives:

\bigskip\noindent
\par\noindent\Enfittab{\begin{tabular}{>{\raggedright\arraybackslash}m{\dimexpr 0.333\linewidth-2\tabcolsep}>{\raggedright\arraybackslash}m{\dimexpr 0.333\linewidth-2\tabcolsep}>{\raggedright\arraybackslash}m{\dimexpr 0.333\linewidth-2\tabcolsep}}
\toprule
\# & Objective & Meaning \\
\midrule
① & Uniform access counts & Over long windows, per-slave and per-level egress counts are close; CV $\approx$ 0 \\
② & Adjacent switching & Consecutive accesses frequently change egress port numbers at each level \\
③ & Outer-level priority & The outermost crossbar egress should rotate first and most frequently \\
\bottomrule
\end{tabular}}

\bigskip

The three objectives in Table 4 correspond to three classes of metrics. Objective ① concerns whether access counts are uniform, measured by terminal \texttt{cv\_slave} and per-level prefix-bucket CV \texttt{cv\_prefix\_Li}. Objective ② concerns whether the outermost egress switches between consecutive accesses, measured by \texttt{flip\_L0} counting whether $f_0$ (i.e., $\text{prefix}_0$) changes. Objective ③ concerns whether inner levels continue to distribute traffic when the outer prefix is fixed: for $i \ge 1$, \texttt{cond\_flip\_Li} counts whether $\text{prefix}_i$ switches between consecutive access pairs where $\text{prefix}_{i -1}$ does not change. Together with \texttt{cv\_prefix\_Li}, \texttt{run\_len}, and \texttt{win\_occ}, this determines whether such switching actually reduces intermediate-level hotspots.
\needspace{7\baselineskip}\par\addvspace{12pt plus 3pt minus 2pt}\noindent{\large \textbf{Metric definitions:}}\par\addvspace{5pt}\noindent
\begin{itemize}
\item \textbf{Terminal balance} (\texttt{cv\_slave}): CV computed over access counts in the compact index domain of active slave nodes. For full-domain \texttt{\{4,8,4\}}, \texttt{cv\_slave} equals \texttt{cv\_prefix\_L2}; for full-domain \texttt{\{4,8,4,2\}}, it equals \texttt{cv\_prefix\_L3}.
\item \textbf{Full-domain terminal CV} (\texttt{cv\_slave\_full}): CV computed over the complete slave node domain as expanded by the physical radix. In harvest pruned scenarios, \texttt{cv\_slave} is computed over the compact active-slave domain.
\item \textbf{Per-level balance} (\texttt{cv\_prefix\_Li}): CV computed over access counts in the level-$i$ prefix buckets. Prefix bucket definition is below.
\item \textbf{Prefix flip rate}: \texttt{flip\_L0} records whether consecutive accesses change $f_0$ (i.e., $\text{prefix}_0$); under random traffic the theoretical upper bound is approximately $1 - 1/r_0$. For $i \ge 1$, \texttt{cond\_flip\_Li} counts only whether $\text{prefix}_i$ flips among consecutive access pairs where $\text{prefix}_{i -1}$ does not change. For \texttt{\{4,8,4\}}, there are 32 L1 prefix buckets: 4 L0 egress ports each containing 8 L1 egress ports. \texttt{cond\_flip\_L1} measures L1 transitions within the same L0 group. \texttt{results/*.txt} retains both \texttt{flip\_Li} and \texttt{cond\_Li} columns; when discussing objective ③, this paper uses the inner-level \texttt{cond\_flip\_Li}.
\item \textbf{Consecutive same-port run length} (\texttt{run\_len}): The number of consecutive accesses to the same prefix bucket. Bucket numbering matches \texttt{cv\_prefix\_Li}. \texttt{run\_len \textgreater  1} means the same output port receives requests on consecutive cycles, a structural hotspot indicator at design time.
\item \textbf{Sliding-window peak occupancy} (\texttt{win\_occ}): The maximum concurrent occupancy of any single prefix bucket within a sliding window of length $\prod_{j=0}^{i} r_j$. For linear, tiling, and block-stride structured stimuli, the ideal occupancy is 1. For lattice or random stimuli, finite windows have a uniform-random Poisson lower bound, which is included as a soft objective $J$.
\end{itemize}

\needspace{7\baselineskip}\par\addvspace{12pt plus 3pt minus 2pt}\noindent{\large \textbf{Prefix buckets and global CV}}\par\addvspace{5pt}\noindent

Per-level CV, \texttt{run\_len}, and \texttt{win\_occ} all share the same prefix bucket definition. The level-$l$ prefix is defined as
\begin{equation}
\text{prefix}_l
= (((f_0 \cdot r_1 + f_1) \cdot r_2 + f_2)\cdots r_l + f_l)
= \sum_{k=0}^{l} f_k \prod_{j=k+1}^{l} r_j,
\end{equation}
i.e., the full routing path ID from root to the level-$l$ egress. Level-$l$ CV, \texttt{run\_len}, and \texttt{win\_occ} are all computed per $\text{prefix}_l$.

CV is computed over the bucket domain $[0,\,\prod_{j=0}^{l} r_j)$, with unvisited buckets counted as 0. For full-domain \texttt{\{4,8,4\}}, the per-level bucket counts are L0=4, L1=32, L2=128; for \texttt{\{4,8,4,2\}}, the last level has 256 buckets. The default experiment window is 1 MiB (8192 accesses at 128-byte granularity), which typically does not cover the complete \texttt{raw} domain. Under sparse lattice stimuli, most buckets are unvisited (count 0) but still included in the CV denominator, so \texttt{cv\_prefix\_Li} and \texttt{cv\_slave} remain large and well-defined.

Each row in \texttt{results/*.txt} is a single snapshot. Columns are organized as: (a) global snapshot columns (\texttt{flip\_L0}, \texttt{cond\_flip\_L1}, \dots); (b) for each level \texttt{Li}, a repeating per-level block [\texttt{flip\_Li}, \texttt{cv\_prefix\_Li}, \texttt{run\_max\_Li}, \texttt{run\_len\_Li}, \texttt{win\_occ\_Li}]. The differential-spectrum CV \texttt{diff\_cv\_observed} computes CV over access counts for address differences $\Delta$, in the set of address differences $\Delta$ actually observed during the experiment (unobserved $\Delta$ are not zero-padded, unlike the prefix-bucket CV above, since the full difference spectrum is too large to enumerate). This metric is computed only in the design-time search tool and does not appear in the \texttt{results/*.txt} wide-table.

\needspace{10\baselineskip}
\subsection{Limitations of Random Access and Terminal CV}

Under a full-rank bijection, random access patterns tend to produce similar \texttt{cv\_slave} values across all candidate algorithms. All close to the random-traffic theoretical lower bound, making it hard to detect arithmetic lattice cliffs or distinguish between mapping algorithms. Folding/table and GF(2) affine have similar CV values under random access, but differ significantly under long strides coprime to the slave count or under diagonal sparse-row access patterns.

Over a complete \texttt{raw} block, terminal counts are exactly uniform (\S{}3 Property 3). When selecting an algorithm, \texttt{cv\_prefix\_Li}, \texttt{run\_len}, and \texttt{win\_occ} must be checked in addition to \texttt{cv\_slave}.

Under random access stimuli, \texttt{flip\_L0} is close to the theoretical upper bound; for inner levels, use \texttt{cond\_flip\_Li}. Under arithmetic lattice stimuli, also use \texttt{cv\_prefix\_Li}, \texttt{run\_len}, and \texttt{win\_occ} (\S{}6.2).

\section{Stimulus Selection and Search Objectives}

\needspace{10\baselineskip}
\subsection{Stimulus (Pattern) Selection}

\needspace{24\baselineskip}
\textbf{Table 5} --- Access pattern priorities:

\bigskip\noindent
\par\noindent\Enfittab{\begin{tabular}{>{\raggedright\arraybackslash}m{\dimexpr 0.333\linewidth-2\tabcolsep}>{\raggedright\arraybackslash}m{\dimexpr 0.333\linewidth-2\tabcolsep}>{\raggedright\arraybackslash}m{\dimexpr 0.333\linewidth-2\tabcolsep}}
\toprule
Priority & Type & Description \\
\midrule
\#1 & linear & Sequential scan with fixed minimum stride; simulates naive burst/stream access \\
\#2 & slice / deslice & Treat contiguous storage as an $M \times N$ matrix; access an $m \times n$ column slice (column-major or reverse column-major, $n < N$, $m < M$); simulates a 1D projection of tiled storage \\
\#3 & block-stride & Contiguous within a block, large-stride jumps between blocks; intra-block accesses are single-element or power-of-two; inter-block address jumps are all powers of two \\
\#4 & matrix sparse-row / 2D lattice & Three lattice projections on a 4096$\times$8192 matrix: row-head 8 columns only, diagonal 8 columns, or block-shift 8 columns; LLM-inspired witnesses (see note) \\
\#5 & non-$2^n$ stride & Strides that are not powers of two, e.g., \texttt{(N-1)$\times$line\_bytes}; coprime to slave count $N$; lowest priority \\
bypass & random & Test only: confirms that flip under a full-rank permutation is close to the theoretical upper bound; not included in search optimization objectives \\
\bottomrule
\end{tabular}}

\bigskip

\textbf{Note (LLM-inspired witnesses):} Geometric access patterns from grouped-query head sharing, paged KV gathering, and dilated attention diagonal strides are drawn from the address patterns of FlashAttention\textsuperscript{[12]}, PagedAttention\textsuperscript{[13]}, blocked GEMM\textsuperscript{[11]}, and tensor-parallel sharding\textsuperscript{[14]}. They are intended to expose arithmetic lattice cliff problems rather than to serve as real production system traces. The nine operator-level witnesses span \#1--\#4 geometries (Table 6); the Priority column gives the threshold bucket (\#1--\#4) each is held to. The \#4-geometry witnesses are reported in \S{}8.1.1 (Table 13) together with the synthetic matrix lattice projections (row-head 8, block-shift 8).

\needspace{30\baselineskip}
\textbf{Table 6} --- Operator-level LLM-inspired witnesses:

\bigskip\noindent
\par\noindent\Enfittab{\begin{tabular}{@{} >{\raggedright\arraybackslash}p{\dimexpr 0.10\linewidth-2\tabcolsep}>{\raggedright\arraybackslash}p{\dimexpr 0.62\linewidth-2\tabcolsep}>{\raggedright\arraybackslash}p{\dimexpr 0.28\linewidth-2\tabcolsep} @{}}
\toprule
Priority & Operator-level witness (benchmark name) & Geometry summary \\
\midrule
\#1 linear & \texttt{ai\_kv\_decode\_4096h8} / \texttt{\_8192h8} & Autoregressive KV row-head append (128 B/step) \\
\#1 linear & \texttt{ai\_tp8\_sharded\_stride128} & TP$\times$8 sharding, 128 B stride sequential stream \\
\#2 slice & \texttt{ai\_batch8\_prefill\_512x4096h8} & batch$\times$seq$\times$hidden prefill tiling \\
\#3 block & \texttt{ai\_gemm\_\{ak,bk,cn\}\_panel\_512x512\_t64} & GEMM A/B/C tile panels \\
\#3 block & \texttt{ai\_weight\_blocked\_512x512\_b64} & Blocked weights 64$\times$64 tile \\
\#3 block & \texttt{ai\_attn\_qk\_kmajor\_128q\_512k\_4096h8} & K-major scan for $Q@K^{\mathsf{T}}$ \\
\#4 lattice & \texttt{diag8\_4096x8192} & 4096$\times$8192 diagonal sparse 8 columns \\
\#4 LLM & \texttt{paged\_kv} / \texttt{attn\_diag16} & Paged KV gather, dilated attention diagonal stride 16 \\
\#4 LLM & \texttt{g8} (grouped query head sharing) & 8-head grouped shared KV \\
\bottomrule
\end{tabular}}
\medskip
\begin{quote}
Full code names are in Table 12.
\end{quote}
\clearpage
\subsection{Search Objectives and Hard Constraints}

\needspace{7\baselineskip}\par\addvspace{12pt plus 3pt minus 2pt}\noindent{\large \textbf{Hard threshold definitions}}\par\addvspace{5pt}\noindent

\textbf{Table 7} lists the admissibility hard thresholds used for design-time parameter selection of \texttt{affine\_pri} (128-node topology, 1 MiB / 8192 access window). Structural stimuli constrain both terminal CV and L0/L1 run/win; lattice stimuli use terminal CV as the hard constraint, with per-level run/win used for diagnostics and soft objectives. Comparison algorithms are included for cross-comparison only and are not filtered by these thresholds.

\needspace{22\baselineskip}
\textbf{Table 7} --- Admissibility hard thresholds:

\bigskip\noindent
\par\noindent\Enfittab{\begin{tabular}{@{}>{\raggedright\arraybackslash}p{\dimexpr 0.17\linewidth-2\tabcolsep}>{\raggedright\arraybackslash}p{\dimexpr 0.14\linewidth-2\tabcolsep}>{\raggedright\arraybackslash}p{\dimexpr 0.08\linewidth-2\tabcolsep}>{\raggedright\arraybackslash}p{\dimexpr 0.24\linewidth-2\tabcolsep}>{\raggedright\arraybackslash}p{\dimexpr 0.31\linewidth-2\tabcolsep}@{}}
\toprule
Check set & Metric & Threshold & Offline verifier & Notes \\
\midrule
\#1--\#3~structural stimuli & \texttt{cv\_slave} & $\leq$ 0.004 & \texttt{K\_CV\_EPS} & Linear/slice/block-stride; close to 0 \\
\#1--\#3~structural stimuli & \texttt{win\_max} (L0, L1) & $\leq$ 2 & \texttt{win\_max $\leq$ WIN\_RUN\_MAX (=2)} & Structural baseline is 1; absolute peak no more than 2 \\
\#1--\#3~structural stimuli & \texttt{run\_max} (L0, L1) & $\leq$ 2 & \texttt{run\_max $\leq$ WIN\_RUN\_MAX (=2)} & Consecutive same-port run no more than 2 \\
\#4 lattice stimuli & \texttt{cv\_slave} & $\leq$ 0.05 & \texttt{K\_LATTICE\_EPS} & Relaxed threshold for arithmetic lattice; 0.005 passes, 0.136 fails (\S{}8.2) \\
Hard constraint for all & Zero slave visits & = 0 & Offline \texttt{make stride127} long-window check & Design-time subsets \#1--\#4 require \texttt{cv\_slave} below threshold; \textbf{zero starving slaves} is verified separately by \textbf{\texttt{make stride127}} long-window check (output {\scriptsize\texttt{STRIDE127\_LONGWINDOW.txt}}; distinct from the stride-127 CV in Table 17 --- from \texttt{run\_hash\_tests} \texttt{linear 16256 B} row, these are two separate paths), not inferred from CV threshold alone \\
\bottomrule
\end{tabular}}

\bigskip

\begin{quote}
Thresholds are checked offline by \texttt{tools/verify\_winner\_prefix\_cv.py} (CV/run/win) and \texttt{tools/stride127\_bench.cc} via \texttt{make stride127} (zero-slave); the C++ benchmark only measures and emits metrics. The \#4 lattice-CV check set is \texttt{\{matrix4096x4096\_row\_diag8\} ∪ AllLatticeConstraintPatternSpecs} (the attention/KV/im2col family selected by \texttt{IsLatticeConstraintPattern()}); \texttt{matrix4096x4096\_row\_diag8} is added via a separate field in \texttt{BuildSearchCorpus} (Appendix B.2), not through \texttt{IsLatticeConstraintPattern()}.
\end{quote}

Sliding window sizes are computed from the topology: level-$l$ window $= \prod_{j=0}^{l} r_j$ (for \texttt{\{4,8,4\}}: L0=4, L1=32, L2=128). Design-time selection uses a representative subset of patterns from \S{}6.1 for fast filtering. The evaluation set in \S{}8 covers more slice, block, matrix, and inference patterns to report generalization behavior and boundaries. The filtering pattern list is in Appendix B.2.

Selection follows a \textbf{hard-constraint + soft-objective} scheme: infeasible $(A, \text{seed})$ pairs are first eliminated per Table 7, and objective function $J$ is then minimized within the feasible set (primarily optimizing \#5 and non-$2^n$ strides). Improving low-priority metrics must not come at the expense of \#1--\#4 hard constraints. Metrics are defined in \S{}5.

Objective function (schematic):
\begin{align}
J &= w_1 \sum_l \mathrm{CV}_l + w_2 \sum_l \alpha_l \mathrm{RunPenalty}_l \nonumber \\
&\quad + w_3 \sum_l \beta_l \mathrm{WindowExcess}_l + w_4 \mathrm{DiffSpectrumPenalty} + w_5 \mathrm{Cost}
\end{align}

Actual weights used in the experiments ($J = J_{cv} + 0.5 J_{win} + 0.2 J_{run} + 0.3 J_{diff}$; this is the instantiated form of the schematic $J$ above, with $w_1\mathrm{CV}_l \to J_{cv}$, $w_2\alpha_l\mathrm{RunPenalty}_l \to J_{run}$, $w_3\beta_l\mathrm{WindowExcess}_l \to J_{win}$, $w_4\mathrm{DiffSpectrumPenalty} \to J_{diff}$; $\alpha_l, \beta_l$ are per-level scaling factors set to 1 in the current implementation; $w_5 \mathrm{Cost}$ has weight 0 and is not included in parameter selection).

Limits for low-priority non-$2^n$ strides:

\begin{itemize}
\item Stride $N -1$ (coprime to slave count $N$) is a known degradation scenario: long-period strides coprime to $N = 2^W$ tend to expose aliasing.
\item In practice, higher CV is tolerated under \#5 patterns than under \#1--\#4, but no slave node may go unvisited for extended periods.
\end{itemize}

\section{Experimental Design}

This section describes the evaluation configuration and comparison algorithms. Artifact build instructions, commands, and result paths are in \S{}7.4.

\needspace{10\baselineskip}
\subsection{Experimental Setup}

C++ benchmark settings: 48-bit logical/encoded addresses (canonical model of \S{}4.1). Line size 128 B (\texttt{L = 7}). \texttt{low} occupies the low 7 bits, \texttt{raw} occupies \texttt{W} bits, and \texttt{high} occupies the remaining high-order address bits. The default statistics window is 1 MiB / 8192 accesses; for pruned topologies, the access count is set to an integer multiple of $N_\text{active}$.

\begin{quote}
\textbf{Implementation note:} The implementation file for 128 slave nodes (\texttt{\{4,8,4\}}) (\texttt{Profile128()}) uses a 36-bit scattered layout (\texttt{scattered\_layout=true}), where idx is distributed across non-contiguous bit fields of the encoded word, differing slightly from the canonical layout of \S{}4.1. The 256-node and harvest topologies use a 48-bit canonical layout. Both layouts are evaluated under the same logical patterns and the same statistics window. Frozen results for the 128-node topology are in \texttt{results/topo484/}. Test topologies are listed in Table 8.
\end{quote}

\needspace{20\baselineskip}
\textbf{Table 8} --- Evaluated topologies:

\bigskip\noindent
\par\noindent\Enfittab{\begin{tabular}{>{\raggedright\arraybackslash}m{\dimexpr 0.200\linewidth-2\tabcolsep}>{\raggedright\arraybackslash}m{\dimexpr 0.200\linewidth-2\tabcolsep}>{\raggedright\arraybackslash}m{\dimexpr 0.200\linewidth-2\tabcolsep}>{\raggedright\arraybackslash}m{\dimexpr 0.200\linewidth-2\tabcolsep}>{\raggedright\arraybackslash}m{\dimexpr 0.200\linewidth-2\tabcolsep}}
\toprule
Topology & Radix & Slave count & $W$ & Role \\
\midrule
Full-domain 3-level & \texttt{\{4,8,4\}} & 128 & 7 & Main text primary \\
Full-domain 4-level & \texttt{\{4,8,4,2\}} & 256 & 8 & Main text (parameterized) \\
Symmetric pruned & \texttt{\{4,7,4\}} & 112 & --- & \S{}8.3 (pruned comparison) \\
Symmetric pruned & \texttt{\{4,7,4,2\}} & 224 & --- & \S{}8.4 (pruned comparison) \\
\bottomrule
\end{tabular}}

\bigskip

\clearpage\subsection{Comparison Algorithms}

\needspace{22\baselineskip}
\textbf{Table 9} --- Comparison algorithms in the main text:

\bigskip\noindent
\par\noindent\Enfittab{\begin{tabular}{>{\raggedright\arraybackslash}m{\dimexpr 0.333\linewidth-2\tabcolsep}>{\raggedright\arraybackslash}m{\dimexpr 0.333\linewidth-2\tabcolsep}>{\raggedright\arraybackslash}m{\dimexpr 0.333\linewidth-2\tabcolsep}}
\toprule
Name & Type & Description and role \\
\midrule
\texttt{low\_order} & Plain bit-slice & The \texttt{raw} address is directly bit-sliced with no permutation. Terminal counts are uniform under most stimuli, but outer-level \texttt{run\_max\_L0} can reach 32 (vs. 2 for \texttt{affine\_pri}). It is used to expose per-level structural hotspots in the absence of any permutation. \\
\texttt{generic\_xor} & Fixed XOR, no search & Simplified XOR bank hash baseline (simplified variant of the linear map framework of\textsuperscript{[2]}): \texttt{idx = raw ⊕ high-bits truncated}, no salt. \\
\texttt{generic\_affine} & GF(2) affine, no search & Full-rank affine baseline, not searched. Same family and same wrapper as \texttt{affine\_pri}; used to isolate the effect of search (same family, with vs. without search). \\
\texttt{affine\_pri} & GF(2) affine, this paper & Matrix and seed frozen by priority-constrained offline search (\S{}4.5). \\
\texttt{type8} & Approach A folding/table & The best internal design from earlier in the project, manually tuned along the folding/table path for this topology. \texttt{cv\_slave} $\approx$ 0 for \#1--\#3, but shows a cliff under arithmetic lattice stimuli (\S{}8.1). \\
\bottomrule
\end{tabular}}

\bigskip

Only the five algorithm types above are expanded in the main text (four baselines + proposed method). Intermediate variants such as fold, gf2fold, and type5 are listed only in the complete results files (\S{}7.4).

\needspace{10\baselineskip}
\subsection{Validation Criteria}

\textbf{Correctness:} At startup, all participating algorithms undergo an injectivity self-check; algorithms with an explicit \texttt{Decode*} also undergo encode/decode round-trip checks. Failures are excluded from results tables.

\needspace{7\baselineskip}\par\addvspace{12pt plus 3pt minus 2pt}\noindent{\large \textbf{Performance metrics:}}\par\addvspace{5pt}\noindent

\begin{itemize}
\item Primary: \texttt{cv\_slave}, \texttt{cv\_prefix\_Li}, and zero-access slave count for low-priority strides.
\item Secondary: \texttt{flip\_L0} and \texttt{cond\_flip\_Li} under random access stimuli; long-window effective bandwidth utilization \texttt{eff\_bw}. Design-time search additionally uses differential-spectrum CV \texttt{diff\_cv\_observed} (see \S{}5; not in benchmark wide-table columns).
\item CV under random access: used as a sanity check only; not included in hard constraint filtering.
\end{itemize}

\textbf{Witness coverage:} The nine stimuli of \S{}6.1 span linear/slice, power-of-two, coprime (\#5), and 2D lattice (\#4) equivalence classes (details in \S{}6.1).

\needspace{10\baselineskip}
\subsection{Artifact and Reproducibility}

\needspace{7\baselineskip}\par\addvspace{12pt plus 3pt minus 2pt}\noindent{\large \textbf{Code and data availability}}\par\addvspace{5pt}\noindent

The benchmark and frozen algorithm implementations for this paper (\texttt{common/}, \texttt{algorithms/}, \texttt{run\_hash\_tests}, figure scripts, and reference \texttt{results/} output) are publicly available at:

\begin{quote}
\href{https://github.com/xiaotongyuan/hsrai\_address\_hash}{https://github.com/xiaotongyuan/hsrai\_address\_hash}\newline
Frozen evaluation tree: \textbf{tag \texttt{paper-v11}} (commit \texttt{adf0f3f}).
\end{quote}

\textbf{Scope of public artifact:} Delivers the code and frozen results needed for evaluation reproducibility (\texttt{make run} / \texttt{make stride127} / Figures 0--3). The design-time parameter search code is not included. Frozen matrices and seeds are in Appendix B; the runtime implementation files such as \texttt{hsrai\_affine\_pri.cc} contain these as constants.

\needspace{7\baselineskip}\par\addvspace{12pt plus 3pt minus 2pt}\noindent{\large \textbf{Build and run (evaluation)}}\par\addvspace{5pt}\noindent

\par{\leftskip=0pt\relax\noindent\hspace*{2em}\begin{minipage}{\linewidth}
\begin{verbatim}
git clone https://github.com/xiaotongyuan/hsrai_address_hash.git
cd hsrai_address_hash
git checkout paper-v11
make run            # 128 slave nodes, all algorithms -> results/topo484/
make run4842        # 256 slave nodes
make run474         # pruned topologies 112/224
make stride127      # stride-127 long window -> results/topo484/STRIDE127_LONGWINDOW.txt
make figures        # Figures 0-3 -> tools/fig*.png / fig*.pdf
\end{verbatim}
\end{minipage}\par}

\needspace{7\baselineskip}\par\addvspace{12pt plus 3pt minus 2pt}\noindent{\large \textbf{Key result files}}\par\addvspace{5pt}\noindent

\begin{itemize}
\item \texttt{results/topo484/SUMMARY.txt}: 128-node topology summary
\item \texttt{results/topo484/hsrai\_affine\_pri.txt}: detailed results for affine\_pri
\end{itemize}

Default evaluation window: 1 MiB / 8192 accesses; pruned topologies use integer-multiple-of-$N_\text{active}$ windows. Stride-127 zero-access checks: \texttt{make stride127} \ensuremath{\rightarrow} \texttt{STRIDE127\_LONGWINDOW.txt}; Table 17 stride-127 CV from full \texttt{run\_hash\_tests} (\texttt{linear}, 16256 B). Figures 0--3: \texttt{make figures}.

\needspace{28\baselineskip}
\textbf{Table 10} --- Algorithm-to-result-file mapping (128 / 112 / 224 node topologies):

\bigskip\noindent
\par\noindent\Enfittab{\begin{tabular}{@{} >{\raggedright\arraybackslash}p{\dimexpr 0.20\linewidth-2\tabcolsep}>{\raggedright\arraybackslash}p{\dimexpr 0.30\linewidth-2\tabcolsep}>{\raggedright\arraybackslash}p{\dimexpr 0.46\linewidth-2\tabcolsep} @{}}
\toprule
Paper name & Code runner name & Main result file \\
\midrule
\texttt{low\_order} & baseline\_low\_order & \texttt{results/topo484/baseline\_low\_order.txt} \\
\texttt{generic\_xor} & baseline\_xor & \texttt{results/topo484/baseline\_xor.txt} \\
\texttt{generic\_affine} & hsrai\_affine128 & \texttt{results/topo484/hsrai\_affine128.txt} \\
\texttt{affine\_pri} & hsrai\_affine\_pri & \texttt{results/topo484/hsrai\_affine\_pri.txt} \\
\texttt{type8} & param\_type8\_scattered & \texttt{results/topo484/param\_type8\_scattered.txt} \\
\texttt{hsrai\_crt474} & hsrai\_crt474 & \texttt{results/topo474/hsrai\_crt474.txt} \\
\texttt{hsrai\_remap} (\S{}4) / \texttt{hsrai\_cyclewalk474} (\S{}8.3) & hsrai\_cyclewalk474 & \texttt{results/topo474/hsrai\_cyclewalk474.txt} \\
\texttt{hsrai\_remap} (\S{}4) / \texttt{hsrai\_cyclewalk4742} (\S{}8.4) & hsrai\_cyclewalk4742 & \texttt{results/topo4742/hsrai\_cyclewalk4742.txt} \\
\bottomrule
\end{tabular}}

\bigskip

\needspace{16\baselineskip}
\textbf{Table 11} --- Algorithm-to-result-file mapping (256 nodes, \texttt{\{4,8,4,2\}}):

\bigskip\noindent
\par\noindent\Enfittab{\begin{tabular}{@{} >{\raggedright\arraybackslash}p{\dimexpr 0.20\linewidth-2\tabcolsep}>{\raggedright\arraybackslash}p{\dimexpr 0.30\linewidth-2\tabcolsep}>{\raggedright\arraybackslash}p{\dimexpr 0.46\linewidth-2\tabcolsep} @{}}
\toprule
Paper name & Code runner name & Main result file \\
\midrule
\texttt{generic\_affine} (256 baseline) & hsrai\_affine\_baseline\_4842 & \texttt{results/topo4842/hsrai\_affine\_baseline\_4842.txt} \\
\texttt{affine\_pri} (256) & hsrai\_affine\_pri\_4842 & \texttt{results/topo4842/hsrai\_affine\_pri\_4842.txt} \\
\bottomrule
\end{tabular}}

\bigskip

\needspace{26\baselineskip}
\textbf{Table 12} --- Pattern name mapping (\S{}6 shorthand \ensuremath{\leftrightarrow} code names):

\bigskip\noindent
\par\noindent\Enfittab{\begin{tabular}{>{\raggedright\arraybackslash}m{\dimexpr 0.500\linewidth-2\tabcolsep}>{\raggedright\arraybackslash}m{\dimexpr 0.500\linewidth-2\tabcolsep}}
\toprule
\S{}6 shorthand & Benchmark stimulus name (in code) \\
\midrule
Row-head 8 columns & \texttt{matrix4096x8192\_row\_head8} \\
diag8 / diagonal 8 columns & \texttt{matrix4096x8192\_row\_diag8} \\
Block-shift sparse row & \texttt{matrix4096x8192\_row\_block\_shift8} \\
g8 & \texttt{ai\_gqa\_kv\_strided\_g8} \\
Paged KV & \texttt{ai\_paged\_kv\_gather\_p128} \\
Attn diag 16 & \texttt{ai\_attn\_diag\_stride16} \\
im2col & \texttt{ai\_conv\_im2col\_s1x1} / \texttt{ai\_conv\_im2col\_s2x2} \\
\bottomrule
\end{tabular}}

\bigskip

\begin{quote}
\textbf{Note:} The stimulus written as \texttt{diag8\_4096x8192} in \S{}6.1 is \texttt{matrix4096x8192\_row\_diag8} in code. The type8 curve in Figure 1 matches the metrics of \texttt{param\_type8\_scattered} (runner name). Naming convention follows \texttt{run\_hash\_tests.cc}.
\end{quote}

\section{Experimental Results}

\needspace{10\baselineskip}
\subsection{Full-Domain 128 Slave Nodes \{4,8,4\}}

All five algorithms have \texttt{cv\_slave} = 0 under \#1--\#3 structural stimuli. This section uses a broader benchmark set than design-time filtering; \texttt{affine\_pri} keeps terminal counts uniform and has no zero-access slaves in the stride-127 long window (\S{}8.1.2). \texttt{low\_order} also shows terminal CV $\approx$ 0 but outer-level run/flip anomalies (\texttt{run\_max\_L0 = 32} vs. 2 for \texttt{affine\_pri}).

Below we focus on \#4 arithmetic lattice / LLM-inspired witnesses and \#5 low-priority patterns. Figure 1 summarizes \texttt{cv\_slave} across \#4 witnesses; \texttt{type8} peaks on lattice patterns while \texttt{affine\_pri} stays low.

\includegraphics[width=\linewidth]{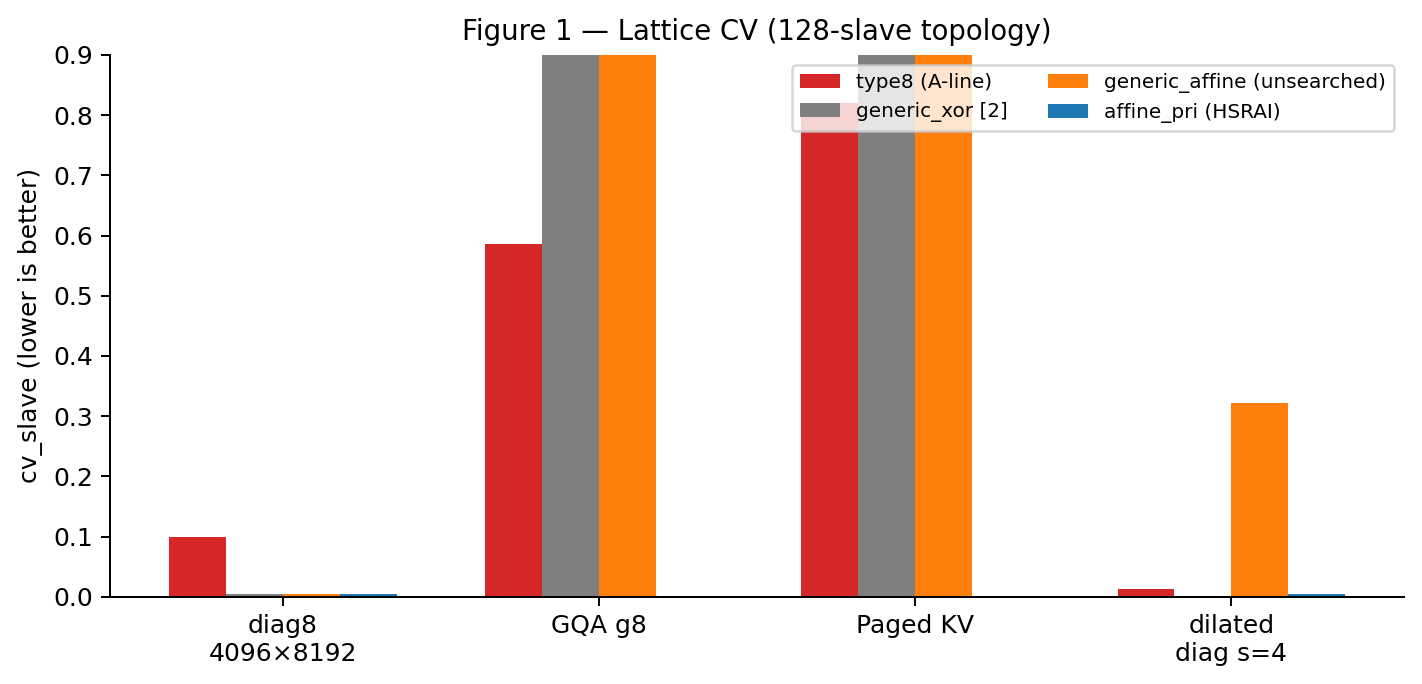}

\clearpage\subsubsection{Arithmetic Lattice and LLM-Inspired Witness Stimuli}

\needspace{24\baselineskip}
\textbf{Table 13} --- \texttt{cv\_slave} for each \#4 witness:

\bigskip\noindent
\par\noindent\Enfittab{\begin{tabular}{>{\raggedright\arraybackslash}m{\dimexpr 0.200\linewidth-2\tabcolsep}>{\raggedright\arraybackslash}m{\dimexpr 0.200\linewidth-2\tabcolsep}>{\raggedright\arraybackslash}m{\dimexpr 0.200\linewidth-2\tabcolsep}>{\raggedright\arraybackslash}m{\dimexpr 0.200\linewidth-2\tabcolsep}>{\raggedright\arraybackslash}m{\dimexpr 0.200\linewidth-2\tabcolsep}}
\toprule
Test & type8 (internal) & generic\_xor\textsuperscript{[2]} & generic\_affine & affine\_pri \\
\midrule
Row-head 8 columns (sequential) & 0 & 2.646 & 0 & 0 \\
Diagonal sparse 8 columns (diag8) & 0.099 & 0.004 & 0.005 & 0.005 \\
Block-shift sparse row & 0.018 & 0 & 0 & 0 \\
Grouped query head sharing (g8) & 0.586 & 1.732 & 1.000 & 0 \\
Paged KV gather & 0.820 & 2.646 & 1.000 & 0 \\
Dilated attention diagonal stride 16 & 0.205 & 0 & 0 & 0 \\
\bottomrule
\end{tabular}}

\bigskip

\texttt{type8} shows a significant cliff when the folding/table period resonates with the lattice stride (g8: 0.586, paged\_kv: 0.820).

\texttt{generic\_xor} does not fail on every \#4 witness: under denser projections like diag8, its terminal \texttt{cv\_slave} is relatively low. The main problems appear in two cases: First, for sparse projections like g8 and paged\_kv where the accesses do not cover all 128 slave nodes, unvisited slave nodes are counted as 0 in \texttt{cv\_slave}, pushing terminal CV to 1.7--2.6. Second, even when terminal CV is low, per-level metrics can still expose intermediate hotspots: for diag8, \texttt{flip\_L0 $\approx$ 0.15} and \texttt{win\_max\_L1 = 17}.

\needspace{14\baselineskip}
\textbf{Table 14} --- Same-family stride-4 comparison (\texttt{cv\_slave}):

\bigskip\noindent
\par\noindent\Enfittab{\begin{tabular}{>{\raggedright\arraybackslash}m{\dimexpr 0.250\linewidth-2\tabcolsep}>{\raggedright\arraybackslash}m{\dimexpr 0.250\linewidth-2\tabcolsep}>{\raggedright\arraybackslash}m{\dimexpr 0.250\linewidth-2\tabcolsep}>{\raggedright\arraybackslash}m{\dimexpr 0.250\linewidth-2\tabcolsep}}
\toprule
Test & generic\_xor\textsuperscript{[2]} & generic\_affine (unsearched) & affine\_pri \\
\midrule
Dilated attention diagonal stride 4 & 0 & 0.322 & 0.004 \\
\bottomrule
\end{tabular}}

\bigskip

\texttt{generic\_affine} and \texttt{affine\_pri} are from the same family with the same wrapper, but the former is unsearched and retains a \texttt{cv\_slave} of 0.322 under stride-4 access. \texttt{affine\_pri}, after priority-constrained search, reduces this to 0.004. The gap is due to the offline search, not the affine family itself.

\needspace{20\baselineskip}
\textbf{Table 15} --- diag8 4096$\times$8192 per-level metrics (\texttt{matrix4096x8192\_row\_diag8}, \texttt{results/topo484/}):

\bigskip\noindent
\par\noindent\Enfittab{\begin{tabular}{>{\raggedright\arraybackslash}m{\dimexpr 0.167\linewidth-2\tabcolsep}>{\raggedright\arraybackslash}m{\dimexpr 0.167\linewidth-2\tabcolsep}>{\raggedright\arraybackslash}m{\dimexpr 0.167\linewidth-2\tabcolsep}>{\raggedright\arraybackslash}m{\dimexpr 0.167\linewidth-2\tabcolsep}>{\raggedright\arraybackslash}m{\dimexpr 0.167\linewidth-2\tabcolsep}>{\raggedright\arraybackslash}m{\dimexpr 0.167\linewidth-2\tabcolsep}}
\toprule
diag8 4096$\times$8192 & cv\_slave & cv\_prefix\_L0 & run\_max\_L0 & win\_max\_L0 & win\_max\_L1 \\
\midrule
type8 (approach A) & 0.099 & 0.013 & 5 & 4 & 6 \\
generic\_xor\textsuperscript{[2]} & 0.004 & 0.002 & 8 & 4 & 17 \\
generic\_affine (unsearched) & 0.005 & 0 & 3 & 3 & 6 \\
affine\_pri & 0.005 & 0 & 3 & 3 & 4 \\
\bottomrule
\end{tabular}}

\bigskip

Although \texttt{generic\_xor} has low terminal CV, \texttt{win\_max\_L1 = 17} shows a transient intermediate-level hotspot. For diag8, \texttt{affine\_pri} has \texttt{win\_max\_L0 = 3} and \texttt{win\_max\_L1 = 4}, matching \texttt{generic\_affine} at the L0 window occupancy level.

\needspace{16\baselineskip}
\textbf{Table 16} --- stride-4 per-level metrics:

\bigskip\noindent
\par\noindent\Enfittab{\begin{tabular}{@{} >{\raggedright\arraybackslash}m{\dimexpr 0.26\linewidth-2\tabcolsep}>{\centering\arraybackslash}m{\dimexpr 0.148\linewidth-2\tabcolsep}>{\centering\arraybackslash}m{\dimexpr 0.148\linewidth-2\tabcolsep}>{\centering\arraybackslash}m{\dimexpr 0.148\linewidth-2\tabcolsep}>{\centering\arraybackslash}m{\dimexpr 0.148\linewidth-2\tabcolsep}>{\centering\arraybackslash}m{\dimexpr 0.148\linewidth-2\tabcolsep} @{}}
\toprule
ai\_attn\_diag\_stride4 & cv\_slave & cv\_prefix\_L0 & run\_max\_L0 & win\_max\_L0 & win\_max\_L1 \\
\midrule
generic\_affine (unsearched) & 0.322 & 0.004 & 2 & 3 & 6 \\
affine\_pri (after search) & 0.004 & 0 & 2 & 2 & 4 \\
\bottomrule
\end{tabular}}

\bigskip

Search reduces L1 window occupancy from 6 to 4 and \texttt{cv\_slave} from 0.322 to 0.004. For the im2col test (\texttt{ai\_conv\_im2col\_s2x2}), both are 0.001.

\needspace{9\baselineskip}
\subsubsection{Low-Priority Access Stimuli}

\#5 low-priority stimuli include non-$2^n$ strides and random access. Figure 2 shows stride-127 starvation (\texttt{type8}: 33/128 zero-access slaves; \texttt{affine\_pri}: none) and \texttt{affine\_pri} effective bandwidth near 1 as the window grows.

\includegraphics[width=\linewidth]{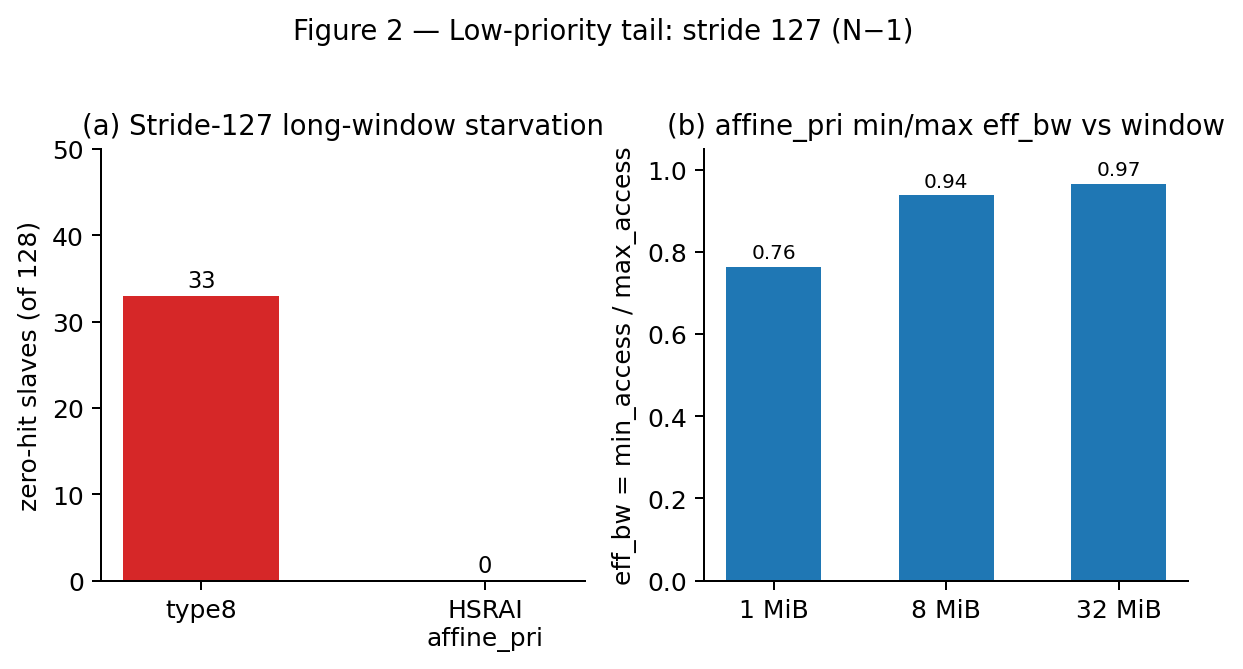}

\needspace{20\baselineskip}
\textbf{Table 17} --- Low-priority access stimulus metrics (stride 127 corresponds to \texttt{linear} pattern, stride 16256 B = 127$\times$128 B line width):

\bigskip\noindent
\par\noindent\Enfittab{\begin{tabular}{>{\raggedright\arraybackslash}m{\dimexpr 0.250\linewidth-2\tabcolsep}>{\raggedright\arraybackslash}m{\dimexpr 0.250\linewidth-2\tabcolsep}>{\raggedright\arraybackslash}m{\dimexpr 0.250\linewidth-2\tabcolsep}>{\raggedright\arraybackslash}m{\dimexpr 0.250\linewidth-2\tabcolsep}}
\toprule
Metric & type8 (internal) & generic\_xor\textsuperscript{[2]} & HSRAI \texttt{affine\_pri} \\
\midrule
Stride-127 CV (\texttt{linear} 16256 B) & 0.847 & 0.125 & 0.062 \\
Worst non-$2^n$ stride CV (\#5) & 0.847@127 & 0.244@384 B & 0.168@10112 B \\
Random access terminal CV & 0.109 & 0.133 & 0.122 \\
Random access outermost flip rate \texttt{flip\_L0} & 0.742 & 0.750 & 0.743 ($\approx$0.75 theoretical) \\
\bottomrule
\end{tabular}}

\bigskip

Table 17 reports stride-127 CV, worst non-$2^n$ stride, and random-access metrics; random-access \texttt{flip\_L0} $\approx$ 0.75 for all three full-rank maps.

\needspace{10\baselineskip}
\subsection{Full-Domain 256 Slave Nodes \{4,8,4,2\}}

The 256-node topology (\texttt{W=8}) is used to test how well parameter search scales with topology size. After the first retarget, \texttt{cv\_slave} for \#1--\#3 stimuli remains 0. Under the \#4 diag8 stimulus, \texttt{cv\_slave} degrades to 0.136, above the lattice threshold of \S{}6.2 (\texttt{K\_LATTICE\_EPS} = 0.05). The \#5 low-priority stride 255 (\texttt{linear 32640} B) has \texttt{cv\_slave} = 0.084; see Table 18. The baseline for this section (\texttt{generic\_affine}, code \texttt{AffineBaselineParams}) uses a pure identity matrix + salt\_seed=0, differing from the near-identity matrix + Salt7FromHigh22 used for the 128-node topology.

Figure 3 compares diag8 \texttt{cv\_slave} distributions for W=7 (128 nodes, fully searched and frozen) and W=8 (256 nodes, first low-budget search iteration).

\includegraphics[width=\linewidth]{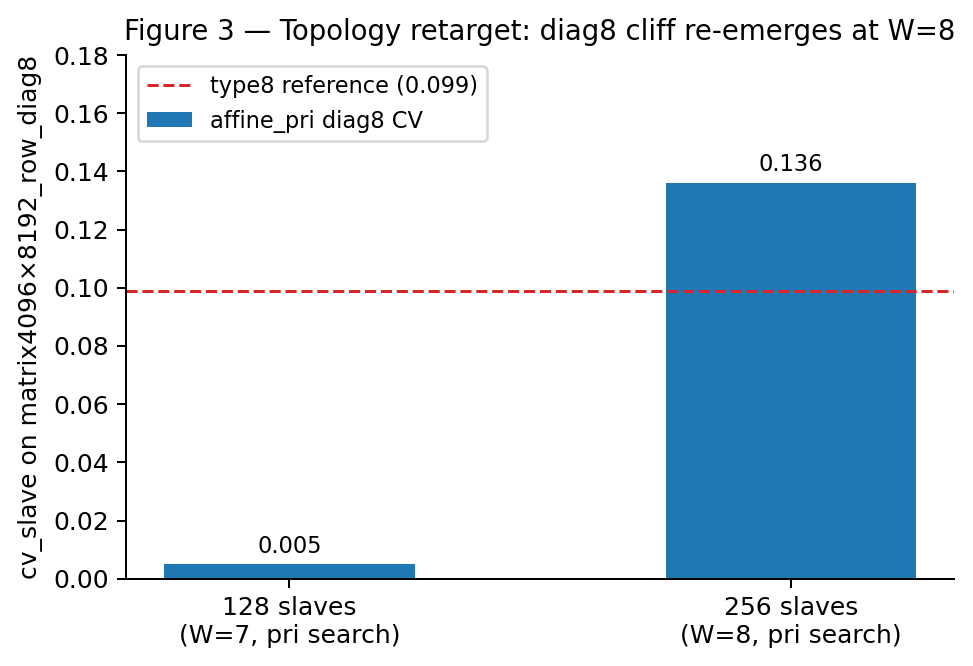}

\needspace{18\baselineskip}
\textbf{Table 18} --- Key observations for the 256-node retarget:

\bigskip\noindent
\par\noindent\Enfittab{\begin{tabular}{>{\raggedright\arraybackslash}m{\dimexpr 0.333\linewidth-2\tabcolsep}>{\raggedright\arraybackslash}m{\dimexpr 0.333\linewidth-2\tabcolsep}>{\raggedright\arraybackslash}m{\dimexpr 0.333\linewidth-2\tabcolsep}}
\toprule
Observation & Primary topology 128 (\texttt{W=7}) & Extended 256 (\texttt{W=8}) \\
\midrule
High-priority \#1--\#3 \texttt{cv\_slave} & = 0 & = 0 \\
\texttt{cv\_slave} for diag8 4096$\times$8192 & 0.005 & 0.136 \\
Low-priority stride 255/127 & 127: 0.062 & 255: 0.084 \\
\bottomrule
\end{tabular}}

\bigskip

\begin{quote}
\textbf{Note:} The first retarget iteration for the 256-node topology used a budget of approximately 300/1500 iterations, well below the budget scale used for the 128-node primary topology. A subsequent expansion to 50,000 iterations in phase C still found no candidate satisfying the \#4 lattice threshold. Table 18 therefore reports the retarget boundary under the current search objectives.
\end{quote}

\needspace{10\baselineskip}
\subsection{Pruned 112 Slave Nodes \{4,7,4\}}

The CRT path (\S{}4.3) decomposes $N_\text{active} = 112 = 2^4 \times 7$ into two sub-domains permuted separately. Tables 19--20 compare \texttt{hsrai\_crt474} against \texttt{gf2fold\_474} (fixed GF(2) fold) and \texttt{hsrai\_cyclewalk474} (remap/cycle-walk).

\needspace{9\baselineskip}
\subsubsection{Key Patterns}

\needspace{28\baselineskip}
\textbf{Table 19} --- \texttt{cv\_slave} for key patterns (112 slave nodes):

\bigskip\noindent
\par\noindent\Enfittab{\begin{tabular}{>{\raggedright\arraybackslash}m{\dimexpr 0.250\linewidth-2\tabcolsep}>{\raggedright\arraybackslash}m{\dimexpr 0.250\linewidth-2\tabcolsep}>{\raggedright\arraybackslash}m{\dimexpr 0.250\linewidth-2\tabcolsep}>{\raggedright\arraybackslash}m{\dimexpr 0.250\linewidth-2\tabcolsep}}
\toprule
Test & gf2fold\_474 & hsrai\_crt474 & hsrai\_cyclewalk474 \\
\midrule
random & 0.101 & 0.079 & 0.089 \\
ai\_gemm\_* panel & 0.000 & 0.000 & 0.000 \\
ai\_kv\_decode\_4096h8 & 0.019 & 0.005 & 0.125 \\
ai\_batch8\_prefill & 0.019 & 0.005 & 0.125 \\
slice\_4096x8192\_w128 & 0.056 & 0.024 & 0.485 \\
diag8 4096$\times$8192 & 0.040 & 0.025 & 0.038 \\
ai\_paged\_kv\_gather & 0.386 & 0.597\textdagger{} & 0.283 \\
ai\_gqa\_kv\_strided\_g8 & 0.399 & 0.302 & 0.358 \\
\bottomrule
\end{tabular}}

\bigskip

\begin{quote}
\textdagger{}: Known degradation scenario for \texttt{hsrai\_crt474}, caused by resonance with odd-factor $q=7$ stride multiples; see note below.
\end{quote}

\texttt{hsrai\_cyclewalk474} has very high residual error under slice access (0.125--0.485) and is not suitable as the primary approach for pruned topologies.

Known degradation: \texttt{paged\_kv\_gather} (0.597) and \texttt{gqa\_kv\_strided\_g8} (0.302) show resonance at stride multiples of the odd factor $q=7$. This is a structural limit of CRT. Deployment must accept this boundary scenario, or switch to remap while accepting its slice degradation.

\needspace{9\baselineskip}
\needspace{20\baselineskip}
\subsubsection{Linear Per-Level Metrics}

\needspace{18\baselineskip}
\textbf{Table 20} --- Linear per-level metrics:

\bigskip\noindent
\par\noindent\Enfittab{\begin{tabular}{>{\raggedright\arraybackslash}m{\dimexpr 0.167\linewidth-2\tabcolsep}>{\raggedright\arraybackslash}m{\dimexpr 0.167\linewidth-2\tabcolsep}>{\raggedright\arraybackslash}m{\dimexpr 0.167\linewidth-2\tabcolsep}>{\raggedright\arraybackslash}m{\dimexpr 0.167\linewidth-2\tabcolsep}>{\raggedright\arraybackslash}m{\dimexpr 0.167\linewidth-2\tabcolsep}>{\raggedright\arraybackslash}m{\dimexpr 0.167\linewidth-2\tabcolsep}}
\toprule
Algorithm & L0 flip & L1 flip & L2 flip & Monotone & run\_max\_L0 \\
\midrule
gf2fold\_474 & 0.042 & 0.224 & 1.000 & No & 49 \\
hsrai\_crt474 & 0.797 & 0.998 & 1.000 & No & 3 \\
hsrai\_cyclewalk474 & 0.077 & 0.186 & 1.000 & No & 44 \\
\bottomrule
\end{tabular}}

\bigskip

The flip columns are per-level \texttt{flip\_Li} from the result files; \texttt{cond\_flip\_Li} is separately reported in the wide table. \texttt{hsrai\_crt474} reduces \texttt{run\_max\_L0} from 49 (\texttt{gf2fold}) to 3, showing that the CRT construction substantially improves outer-level consecutive same-port run length under linear access.

\needspace{22\baselineskip}
\subsection{Pruned 224 Slave Nodes \{4,7,4,2\}}

$N_\text{active} = 224 = 2^5 \times 7$. The CRT path is similar to the 112-node topology, extended to 4 levels. Tables 21--22 compare \texttt{hsrai\_crt4742} against \texttt{gf2fold\_4742} (fixed GF(2) fold) and \texttt{hsrai\_cyclewalk4742} (remap/cycle-walk).

\needspace{9\baselineskip}
\subsubsection{Key Patterns}

\needspace{28\baselineskip}
\textbf{Table 21} --- \texttt{cv\_slave} for key patterns (224 slave nodes):

\bigskip\noindent
\par\noindent\Enfittab{\begin{tabular}{>{\raggedright\arraybackslash}m{\dimexpr 0.250\linewidth-2\tabcolsep}>{\raggedright\arraybackslash}m{\dimexpr 0.250\linewidth-2\tabcolsep}>{\raggedright\arraybackslash}m{\dimexpr 0.250\linewidth-2\tabcolsep}>{\raggedright\arraybackslash}m{\dimexpr 0.250\linewidth-2\tabcolsep}}
\toprule
Test & gf2fold\_4742 & hsrai\_crt4742 & hsrai\_cyclewalk4742 \\
\midrule
random & 0.129 & 0.127 & 0.133 \\
ai\_gemm\_* panel & 0.000 & 0.000 & 0.000 \\
ai\_kv\_decode\_4096h8 & 0.025 & 0.008 & 0.216 \\
ai\_batch8\_prefill & 0.025 & 0.008 & 0.216 \\
slice\_4096$\times$8192\_w128 & 0.069 & 0.025 & 0.706 \\
diag8 4096$\times$8192 & 0.076 & 0.082 & 0.089 \\
ai\_paged\_kv\_gather & 0.771 & 0.577 & 0.690 \\
ai\_gqa\_kv\_strided\_g8 & 0.557 & 0.513 & 0.419 \\
\bottomrule
\end{tabular}}

\bigskip

\texttt{hsrai\_cyclewalk4742} has even higher slice residual error (0.216--0.706). Factor-7 resonance in paged\_kv/g8 matches \S{}8.3.

\needspace{9\baselineskip}
\needspace{20\baselineskip}
\subsubsection{Linear Per-Level Metrics}

\needspace{18\baselineskip}
\textbf{Table 22} --- Linear per-level metrics (4 levels):

\bigskip\noindent
\par\noindent\Enfittab{\begin{tabular}{>{\raggedright\arraybackslash}m{\dimexpr 0.167\linewidth-2\tabcolsep}>{\raggedright\arraybackslash}m{\dimexpr 0.167\linewidth-2\tabcolsep}>{\raggedright\arraybackslash}m{\dimexpr 0.167\linewidth-2\tabcolsep}>{\raggedright\arraybackslash}m{\dimexpr 0.167\linewidth-2\tabcolsep}>{\raggedright\arraybackslash}m{\dimexpr 0.167\linewidth-2\tabcolsep}>{\raggedright\arraybackslash}m{\dimexpr 0.167\linewidth-2\tabcolsep}}
\toprule
Algorithm & L0 flip & L1 flip & L2 flip & L3 flip & run\_max\_L0 \\
\midrule
gf2fold\_4742 & 0.021 & 0.110 & 0.429 & 1.000 & 110 \\
hsrai\_crt4742 & 0.710 & 0.785 & 1.000 & 0.684 & 4 \\
hsrai\_cyclewalk4742 & 0.040 & 0.089 & 0.429 & 1.000 & 72 \\
\bottomrule
\end{tabular}}

\bigskip

\texttt{hsrai\_crt4742} reduces \texttt{run\_max\_L0} from 110 (\texttt{gf2fold}) to 4, an improvement consistent with the 112-node topology.

\section{Related Work}

We survey existing work along three threads: XOR/bank hashing; partition camping and tensor access; and non-$2^n$ permutation constructions. Each subsection clarifies the assumptions typically made by flat bank hashing and similar schemes, and identifies where HSRAI extends them with permutation-preserving, per-level acceptance (cf. \S{}1).

\needspace{10\baselineskip}
\subsection{XOR / Bank Hashing}

Seznec \& Espasa\textsuperscript{[1]} give theoretical conditions for stride-conflict-free access in $2^n$ cache banks. Vandierendonck \& De Bosschere\textsuperscript{[2]} model XOR bank hashing as a GF(2) linear map and analyze its conflict-free properties. Seznec skewed cache\textsuperscript{[3]}, Minimalist Open-page\textsuperscript{[4]}, and configurable XOR scratchpad for GPUs\textsuperscript{[5]} show that low-order XOR mixing is widely used in caches, DRAM controllers, and on-chip storage with low hardware cost. All of these works target single-level bank/partition indexing; they neither require a global bijection between logical and encoded addresses nor consider per-level CV and window metrics for multi-level radix crossbars.

The industrial folding/table approach (\S{}2.2 approach A) lacks a peer-reviewed algorithm description that can be directly cited. This paper uses the manually tuned \texttt{type8} as the internal reference implementation for that approach (\S{}7.2); this is not intended as a claim that it represents any published algorithm. HSRAI, while preserving GF(2) full-rank invertibility (\S{}3 Property 2), extends the acceptance target from ``single-level conflict-free'' to static admissibility at each level's routing egress $\{f_i\}$ by searching offline for matrix $A$ and salt.

\begin{quote}
\textbf{Note:} The \texttt{generic\_xor} baseline in the experiments takes the simplified form $\text{idx} = \text{raw} \oplus \text{high -bits truncated}$ (no salt; high bits use only the low $W$ bits for XOR). This is a simplified variant of the linear map framework of\textsuperscript{[2]}, not a complete reimplementation.
\end{quote}

\needspace{10\baselineskip}
\subsection{Partition Camping and Tensor Access Motivation}

MemPool\textsuperscript{[6]} in a shared-L1 multi-core cluster maps some memory regions sequentially and others with interleaving, showing that the address mapping strategy attached to a NoC is a first-order design decision for bandwidth utilization. However, MemPool uses a static partition strategy for memory regions rather than a global approach that applies a single invertible permutation over the entire address space with per-level search optimization; the two approaches address problems at different levels of abstraction. Aji et al.\textsuperscript{[7]} empirically show that regular tensor accesses can leave GPU memory partitions without accesses for extended periods, consistent with the cliff motivation of the \#4 lattice witnesses in \S{}6.1.

FlashAttention\textsuperscript{[12]}, PagedAttention\textsuperscript{[13]}, blocked GEMM\textsuperscript{[11]}, and Megatron-LM\textsuperscript{[14]} provide the geometric basis for the address patterns used in the structured synthetic stimuli of \S{}6.1 (structured synthetic witnesses). Software-side data layout optimization can relieve hotspots in some scenarios, but when the hardware fabric hash is fixed or opaque to software, the interleaver's own balancing quality cannot be compensated by software.

\needspace{10\baselineskip}
\subsection{Non-$2^n$ Domains and Other Permutation Families}

Seznec \& Lenfant\textsuperscript{[8]}, Gao\textsuperscript{[9]}, and Valero et al.\textsuperscript{[10]} explored odd-modulus memory interleaving, the Chinese Remainder Theorem, and modulo skewing respectively; these are the theoretical sources for the CRT path in HSRAI pruned topologies (approach C, \S{}4.3). Full-domain designs do not use prime division to avoid runtime division overhead. The CRC/LFSR scrambling commonly used in industry\textsuperscript{[15]} is typically a many-to-one mapping and does not satisfy the global bijection constraint of \S{}2.1.

Costas arrays\textsuperscript{[16]}, Latin square complete mappings\textsuperscript{[17]}, and distinct-difference constructions\textsuperscript{[16,17]} provide algebraic construction methods for conflict-free access. Feistel/FPE-type finite-domain permutations\textsuperscript{[18,21]} provide another family of bijective constructions. Attaching Costas/Latin constructions as small external tables on top of an affine core makes it hard to simultaneously improve per-level metrics under the \#1--\#3 hard constraints. Deeper algebraic integration is left as future work (Appendix A.2).

\section{Conclusion}

This paper studies permutation-preserving address interleaving on multi-level interconnect fabrics. Beyond terminal slave-node balance, any feasible mapping must be a global bijection and keep each level's egress ports as uniform as possible under structured accesses---acceptance targets stricter than the flat bank hashing common in caches, DRAM, and GPUs\textsuperscript{[1,2,5]}.

HSRAI preserves high-order and intra-line low-order address bits and applies a $W$-bit bijection only to the intermediate index segment. Full-domain topologies use offline GF(2) affine search with priority hard constraints (\texttt{affine\_pri}); pruned topologies combine the Chinese Remainder Theorem\textsuperscript{[8--10]} with optional remapping. On a reproducible structured synthetic benchmark, admissibility is evaluated by per-level CV, run length, and sliding-window occupancy.

On the 128-node primary topology, \texttt{affine\_pri} is the most consistently robust among the evaluated schemes: it meets high-priority balance and long-window non-starvation requirements and remains stable under arithmetic-lattice and LLM-inspired stimuli. Simplified fixed-XOR implementations, unsearched full-rank affine maps, and a manually tuned folding/table reference algorithm each expose weaknesses on a different dimension---intermediate-level hotspots, residual stride contention, or extended zero-access periods. Retarget experiments on the 256-node topology and the CRT path on pruned topologies with an odd factor yield explicitly reported search and structural boundaries.

For designers: first determine workload priorities and set hard-constraint filters; then screen mapping algorithms and parameters with per-level admissibility metrics; finally compare low-priority access-pattern performance within the feasible set. Future directions include joint search over bit mapping and CRT/number-theoretic constructions, deeper algebraic integration of Latin squares and Costas arrays\textsuperscript{[16,17]}, validation on real address traces, and RTL signoff; scope limitations are discussed in Appendix A.

\section*{Declaration of Generative AI and AI-assisted technologies in the writing process}

During the preparation of this work, the author used Cursor (with embedded models from OpenAI, Anthropic and Zhipu AI) to support manuscript organization, English translation and polishing, terminology consistency checks, formatting and cross-reference checks, and drafting of code and documentation fragments under author-provided specifications. After using these tools, the author reviewed and edited the content as needed and takes full responsibility for the content of the published article.

Generative AI tools were not listed as authors and were not treated as independent scientific sources. The problem definition, mathematical formulation, experimental design, interpretation of results, and conclusions were determined and checked by the author. Experimental data, search results, quantitative tables, and figures come from author-run benchmark outputs and the frozen public artifact; reference entries were recorded and verified by the author.

\clearpage\mwmsection*{References}

\textbf{[1]} A. Seznec and R. Espasa, ``Conflict-free accesses to strided vectors on a banked cache,'' \textit{IEEE Trans. Comput.}, vol. 54, no. 7, pp. 913--926, Jul. 2005.

\textbf{[2]} H. Vandierendonck and K. De Bosschere, ``XOR-based hash functions,'' \textit{IEEE Trans. Comput.}, vol. 54, no. 7, pp. 800--812, Jul. 2005.

\textbf{[3]} A. Seznec, ``A case for two-way skewed-associative caches,'' in \textit{Proc. ISCA}, 1993, pp. 169--178.

\textbf{[4]} D. Kaseridis, J. Stuecheli, and L. K. John, ``Minimalist open-page: A DRAM page-mode scheduling policy for the many-core era,'' in \textit{Proc. MICRO}, 2011, pp. 24--35.

\textbf{[5]} G.-J. van den Braak et al., ``Configurable XOR hash functions for banked scratchpad memories in GPUs,'' \textit{IEEE Trans. Comput.}, vol. 65, no. 7, pp. 2045--2058, Jul. 2016.

\textbf{[6]} M. A. Cavalcante et al., ``MemPool: A shared-L1 memory many-core cluster with a low-latency interconnect,'' in \textit{Proc. DATE}, 2021; extended version published in \textit{IEEE Trans. Comput.}, vol. 72, no. 12, 2023.

\textbf{[7]} A. M. Aji, M. Daga, and W.-c. Feng, ``Bounding the effect of partition camping in GPU kernels,'' in \textit{Proc. CF}, 2011.

\textbf{[8]} A. Seznec and J. Lenfant, ``Odd memory systems may be quite interesting,'' in \textit{Proc. ISCA}, 1993, pp. 341--350.

\textbf{[9]} Q. S. Gao, ``The Chinese remainder theorem and the prime memory system,'' in \textit{Proc. ISCA}, 1993, pp. 337--340.

\textbf{[10]} M. Valero et al., ``Increasing the number of strides for conflict-free vector access,'' in \textit{Proc. ISCA}, 1992, pp. 372--381.

\textbf{[11]} V. Volkov and J. W. Demmel, ``Benchmarking GPUs to tune dense linear algebra,'' in \textit{Proc. SC}, 2008.

\textbf{[12]} T. Dao et al., ``FlashAttention: Fast and memory-efficient exact attention with IO-awareness,'' in \textit{Proc. NeurIPS}, 2022.

\textbf{[13]} W. Kwon et al., ``Efficient memory management for large language model serving with PagedAttention,'' in \textit{Proc. SOSP}, 2023.

\textbf{[14]} D. Narayanan et al., ``Efficient large-scale language model training on GPU clusters using Megatron-LM,'' in \textit{Proc. SC}, 2021.

\textbf{[15]} M. Grymel and S. B. Furber, ``A novel programmable parallel CRC circuit,'' \textit{IEEE Trans. Very Large Scale Integr. (VLSI) Syst.}, vol. 19, no. 10, pp. 1898--1902, Oct. 2011.

\textbf{[16]} S. W. Golomb, ``Algebraic constructions for Costas arrays,'' \textit{J. Combin. Theory Ser. A}, vol. 37, no. 1, pp. 13--21, Jul. 1984.

\textbf{[17]} C. J. Colbourn and K. Heinrich, ``Conflict-free access to parallel memories,'' \textit{J. Parallel Distrib. Comput.}, vol. 14, no. 2, pp. 193--200, Feb. 1992.

\textbf{[18]} M. Bellare, T. Ristenpart, P. Rogaway, and T. Stegers, ``Format-preserving encryption,'' in \textit{Proc. Selected Areas in Cryptography (SAC)}, Calgary, AB, Canada, Aug. 2009, pp. 295--312.

\textbf{[19]} N. McKeown, ``The iSLIP scheduling algorithm for input-queued switches,'' \textit{IEEE/ACM Trans. Networking}, vol. 7, no. 2, pp. 188--201, Apr. 1999.

\textbf{[20]} M. Karol, M. Hluchyj, and S. Morgan, ``Input versus output queuing on a space-division packet switch,'' \textit{IEEE Trans. Commun.}, vol. 35, no. 12, pp. 1347--1356, Dec. 1987.

\textbf{[21]} B. Morris, P. Rogaway, and T. Stegers, ``How to encipher messages on a small domain: Deterministic encryption and the Thorp shuffle,'' in \textit{Proc. Advances in Cryptology (CRYPTO)}, Santa Barbara, CA, USA, Aug. 2009, pp. 286--302.

\section*{Appendix A: Discussion and Limitations}

\subsection*{A.1 Experimental Scope Limitations}

The evaluation is based on static access counts and structured synthetic stimuli, with the following scope limitations:

\textbf{No RTL synthesis.} This paper delivers a design-time admissibility framework (a mapping filter), not a cycle-accurate bandwidth model. \texttt{win\_max}/\texttt{run\_len} are design-time hotspot indicators, not proven throughput lower bounds. For example, for diag8 (\texttt{matrix4096x8192\_row\_diag8}), \texttt{generic\_xor} has \texttt{win\_max\_L1 = 17} while \texttt{affine\_pri} has 4 (Table 15), a significant reduction in intermediate-level sliding-window peak occupancy. In virtual output queue (VOQ)\textsuperscript{[20]} interconnects (such as the shared scratchpad interconnect described in MemPool\textsuperscript{[6]}), high peak occupancy is a first-order indicator of egress queue congestion; mapping selection is a prerequisite decision for interconnect design (this paper does not perform queue simulation signoff). Hardware cost analysis is operator-level estimation only and has not been verified by RTL synthesis. \textbf{Table 23} gives a rough comparison for the 128-node topology:

\needspace{18\baselineskip}
\textbf{Table 23} --- Algorithm hardware cost comparison (128-node topology, order-of-magnitude):

\bigskip\noindent
\par\noindent\Enfittab{\begin{tabular}{>{\raggedright\arraybackslash}m{\dimexpr 0.333\linewidth-2\tabcolsep}>{\raggedright\arraybackslash}m{\dimexpr 0.333\linewidth-2\tabcolsep}>{\raggedright\arraybackslash}m{\dimexpr 0.333\linewidth-2\tabcolsep}}
\toprule
Algorithm & Logic structure & Storage requirement \\
\midrule
type8 & Two-level 128-entry permutation tables + folding adder path & Table entries must be refilled when topology changes \\
hsrai\_affine\_pri & 7$\times$7 GF(2) matrix multiply (approximately a few dozen XORs) + salt logic & No large tables; parameters frozen as constants \\
hsrai\_crt474 & Small modular arithmetic combination over the 112-entry domain & No large tables; higher latency than pure XOR but pipelineable \\
\bottomrule
\end{tabular}}

\bigskip

The above are order-of-magnitude estimates showing the trade-off direction between approaches B/C and approach A in terms of parameterizability and area. Precise area/timing/power figures require RTL synthesis (see A.2 future work).

\textbf{Why only single-stream sustained rate, not multi-master transient peaks, is used for algorithm selection.} In a standard VOQ\textsuperscript{[20]} + iSLIP\textsuperscript{[19]} + backpressure interconnect, transient peaks caused by multiple masters aligning in the same cycle are absorbed within a few cycles by backpressure and self-heal (arbitration serializes rather than drops traffic); their cost is reflected in buffer depth and queuing delay, which are system-level parameters. The hard upper bound on throughput is determined by the sustained rate uniformity at each level's egress (admissibility: per-level CV + run-length). Therefore, this paper restricts \textbf{mapping selection} objectives to single-stream per-level CV and run-length; multi-master same-cycle peaks are not included in search objective $J$ (this paper does not perform VOQ queue simulation signoff). This is an intentional scope decision: mapping determines long-term aggregate uniformity (which backpressure cannot compensate for); scheduling policy and buffer depth determine transient timing (which backpressure can compensate for). \texttt{win\_max}/\texttt{run\_len} serve as design-time hotspot indicators under this acceptance rule, not as cycle-accurate throughput proofs.

\textbf{No production traces.} Evaluation uses structured synthetic stimuli covering linear, slice, arithmetic lattice, and LLM-inspired geometries (\S{}6.1). FlashAttention\textsuperscript{[12]}, PagedAttention\textsuperscript{[13]}, and blocked GEMM\textsuperscript{[11]} provide the geometric basis for the address patterns used in the stimuli, but these are not equivalent to real address traces from production systems. The representativeness of the stimulus design requires validation with real traces.

\textbf{Not globally optimal.} \texttt{affine\_pri} is the best engineering point under the current objective function and stimulus set. The 256-node diagonal sparse, stride-255, and other retarget and low-priority stimuli still have room for improvement. The 112-node CRT resonates under paged-KV and factor-7-related stimuli (CV 0.30--0.60; see Table 19). Deployment in harvest topologies must accept this known degradation scenario, switch to remap and accept its slice degradation, or increase $N_\text{active}$ (\S{}8.3--\S{}8.4).

\subsection*{A.2 Future Work}

\textit{Evaluation and comparison}

\begin{itemize}
\item Real address trace validation: Validate the coverage representativeness of structured synthetic stimuli using real traces dumped from accelerators (there is currently no widely adopted public standard benchmark for NoC/fabric address traces).
\item More public comparison schemes: Low-order interleaving and generic XOR bank hash\textsuperscript{[2]} are already included as public baselines (\S{}7.2). Further comparison with more published bank-hash variants\textsuperscript{[1,3--5]} can be done in follow-up work.
\end{itemize}

\textit{Algorithms and search}

\begin{itemize}
\item Breaking the structural ceiling for non-$2^n$ low-priority and 2D lattice patterns: GF(2) affine has a structural ceiling on integer-stride carry behavior (\S{}6.2, Property 4). Future directions include deeper integration of approach C CRT/number-theoretic decomposition\textsuperscript{[8--10]}, and incorporating Latin square skewing\textsuperscript{[17]}, Costas arrays\textsuperscript{[16]}, and distinct-difference constructions\textsuperscript{[16,17]} into the permutation core or CRT construction (rather than attaching them as external small tables).
\item Joint bit-mapping and matrix search: Include the idx\ensuremath{\rightarrow}$\{f_i\}$ bit-mapping together with matrix $A$ in a joint search space. The current set-cover approach uses only a pow2-bit heuristic seed, with limited effect.
\item Carry-aware salt and differential-spectrum guidance: Introduce CRC\textsuperscript{[15]}/carry-proxy family salt (carry-proxy is a search heuristic in this paper) and stride differential-spectrum penalty terms, to detect lattice-sensitive directions early and guide the search.
\item Permutation family extensions: Evaluate Feistel-type format-preserving encryption (FPE)\textsuperscript{[18,21]}, Costas arrays\textsuperscript{[16]}, and Latin square complete mappings\textsuperscript{[17]} against GF(2) affine\textsuperscript{[2]} in terms of per-level CV trade-offs.
\end{itemize}

\textit{Implementation}

\begin{itemize}
\item RTL synthesis and signoff: Perform RTL synthesis of approach B GF(2) XOR networks and approach C CRT small modular arithmetic to obtain area/timing/power data, replacing the current operator-level estimates.
\end{itemize}

\section*{Appendix B: Frozen Parameters}

This appendix lists the frozen parameters for each topology, for cross-validation and reproducibility of the evaluation. GF(2) matrices are given in row-vector form; each byte represents the bit mask for that row (bit 0 corresponds to column 0). Parameters are selected offline at design time and written into the source files listed below. The public artifact uses these constants as the definitive reference. The selection program itself is not included.

Artifact version pinning (\S{}7.4): Repository \texttt{https://github.com/xiaotongyuan/hsrai\_address\_hash}, tag \texttt{paper-v11}, commit \texttt{adf0f3f}.

\subsection*{B.0 Full-Domain 128-Node Baseline Parameters (generic\_affine, \{4,8,4\}, W=7)}

\begin{itemize}
\item \textbf{$A_\text{base}$} (7 rows $\times$ 7 cols, row vectors in hex): \texttt{\{0x01, 0x02, 0x04, 0x08, 0x10, 0x20, 0x47\}} (near-identity; last row \texttt{0x47} is non-trivial)
\item Salt function: \texttt{Salt7FromHigh22} (\S{}4.2); no seed, parameters fixed
\item Source: \texttt{hsrai\_affine128.cc} (\texttt{kA} array); not searched; baseline comparison
\end{itemize}

\needspace{7\baselineskip}\subsection*{B.1 Full-Domain 128 Slave Nodes (affine\_pri, \{4,8,4\}, W=7)}

\begin{itemize}
\item \textbf{$A_\text{pri}$} (7 rows $\times$ 7 cols, row vectors in hex): \texttt{\{0x45, 0x10, 0x68, 0x48, 0x41, 0x16, 0x33\}}
\item salt seed: \texttt{0xF1BB3E62}
\item Salt function: $\texttt{CRC32}(\text{high} \oplus \text{seed}) \bmod 2^7$
\item Source: \texttt{hsrai\_affine\_pri.cc} (selected at design time per \S{}6.2; verified by \S{}8 benchmarks)
\end{itemize}

\subsection*{B.2 Design-Time Filtering Set (128-node topology, \texttt{BuildSearchCorpus})}

The design-time search uses the following representative patterns for hard-threshold filtering (consistent with \texttt{HardConstraintsOk} in \texttt{search\_tools}):

\bigskip\noindent
\par\noindent\Enfittab{\begin{tabular}{>{\raggedright\arraybackslash}m{\dimexpr 0.22\linewidth-2\tabcolsep}>{\raggedright\arraybackslash}m{\dimexpr 0.74\linewidth-2\tabcolsep}}
\toprule
Category & Benchmark stimulus name \\
\midrule
linear N=1 & \texttt{linear} stride 128 B \\
slice (5) & \texttt{slice\_4096x4096\_w512}, \texttt{w2048}; \texttt{slice\_4096x8192\_w512}, \texttt{w2048}; \texttt{slice\_4096x4096\_w4096} \\
block (3) & \texttt{blk\_acc256\_skip256}, \texttt{blk\_acc1024\_skip1024}, \texttt{blk\_acc4096\_skip4096} \\
\#4 lattice (CV) & \texttt{matrix4096x4096\_row\_diag8} and full \texttt{AllLatticeConstraintPatternSpecs} set \\
\bottomrule
\end{tabular}}

\bigskip

The \texttt{results/topo484/} wide table in the public artifact covers all of the above patterns and extends to more slice, block, matrix, and inference rows. Lattice run/win are reported as diagnostic metrics.

\subsection*{B.3 Full-Domain 256-Node Stress Test Parameters (affine\_pri\_4842, \{4,8,4,2\}, W=8)}

\begin{itemize}
\item \textbf{$A_\text{pri}$} (8 rows $\times$ 8 cols, row vectors in hex): \texttt{\{0x4D, 0xB6, 0xB2, 0x16, 0xC8, 0x86, 0x54, 0xAD\}}
\item salt seed: \texttt{0x64F46100}
\item Source: \texttt{hsrai\_affine\_profile.cc} (design-time {\textasciitilde}300/1500 iteration scale; see \S{}8.2)
\item Note: First-iteration search result; \#4 lattice \texttt{cv\_slave} constraint not fully satisfied (see \S{}8.2); included as stress-test reference only
\end{itemize}

\subsection*{B.4 Pruned 112 Slave Nodes (crt474, \{4,7,4\}, $N_\text{active}=112$)}

\begin{itemize}
\item \textbf{2\textsuperscript{4} sub-domain} $A_2$ (4 rows $\times$ 4 cols, row vectors in hex): \texttt{\{0x0B, 0x08, 0x05, 0x04\}}
\item mod 7 sub-domain: $a_7 = 1$, $b_7 = 0$ (i.e., $p_q = (r_q + \text{salt}_q) \bmod 7$, consistent with the general formula of \S{}4.3 at $q=7$; $r_q$ is the mod-7 sub-domain component)
\item salt seed: \texttt{0xB175F377} (shared by both sub-domains; 2\textsuperscript{4} domain truncates to 4 bits; mod-7 domain uses \texttt{seed ⊕ 0xA7A7A7A7} then mod 7)
\item Source: \texttt{hsrai\_crt474.cc} (design-time selection; \texttt{474} and \texttt{harvest112} in code comments are aliases for the same topology)
\end{itemize}

\subsection*{B.5 Pruned 224 Slave Nodes (crt4742, \{4,7,4,2\}, $N_\text{active}=224$)}

\begin{itemize}
\item \textbf{2\textsuperscript{5} sub-domain} $A_2$ (5 rows $\times$ 5 cols, row vectors in hex): \texttt{\{0x1F, 0x0E, 0x1D, 0x1C, 0x0B\}}
\item mod 7 sub-domain: $a_7 = 5$, $b_7 = 1$ (i.e., $p_q = (5 r_q + 1 + \text{salt}_q) \bmod 7$, consistent with \S{}4.3 general formula)
\item salt seed: \texttt{0x5C4D3E00}
\item Source: \texttt{hsrai\_crt474.cc} (design-time selection; \texttt{4742} and \texttt{harvest224} in code comments are aliases for the same topology)
\end{itemize}


\end{document}